\pdfoutput=1
\documentclass[preprint,12pt,authoryear]{elsarticle}
\usepackage{amsmath,amssymb,amsthm,amsfonts}
\usepackage{mathtools}
\usepackage{placeins}
\usepackage{multirow}
\usepackage{booktabs}
\usepackage{subcaption}
\usepackage{tikz}
\usepackage{siunitx}
\usepackage{annotate-equations}
\usepackage{graphicx}
\usepackage{url}
\usepackage[hidelinks]{hyperref}
\usepackage[margin=2cm]{geometry}
\graphicspath{{./}}
\journal{Engineering Applications of Artificial Intelligence}
\makeatletter
\AtBeginDocument{%
  \def\@currentlabel{\ensuremath{\dagger}}\elsLabel{myc}%
  \def\@currentlabel{\ensuremath{\ddagger}}\elsLabel{sm}%
  \def\@currentlabel{\ensuremath{\mathsection}}\elsLabel{icl}}
\makeatother

\newtheorem{property}{Property}
\theoremstyle{definition}
\newtheorem{example}{Example}
\newenvironment{continuedexample}[1]{%
  \par\addvspace{\medskipamount}%
  \noindent\textbf{Example~\ref{#1} (continued)}%
  \hspace{1em}%
  \itshape\ignorespaces
}{\par\addvspace{\medskipamount}}

\newif\ifanon
\ifdefined\ANON\anontrue\fi

\begin{document}
\begin{frontmatter}

\title{A decomposition approach to solving numerical constraint satisfaction problems on directed acyclic graphs}

\ifanon\else
\author[myc,sm,icl]{Max Mowbray}
\ead{max@myceliai.co}
\author[icl]{Nilay Shah}
\ead{n.shah@imperial.ac.uk}
\author[icl]{Beno\^it Chachuat\corref{cor1}}
\ead{b.chachuat@imperial.ac.uk}
\cortext[cor1]{Corresponding author.}
\address{$^{\dagger}$Mycelia Intelligence, London, United Kingdom}
\address{$^{\ddagger}$Secondmind, Cambridge, United Kingdom}
\address{$^{\mathsection}$The Sargent Centre for Process Systems Engineering, Imperial College London, London, United Kingdom}
\fi

\begin{abstract}
Certifying that a decision satisfies every constraint of an engineered system can be framed as a constraint satisfaction problem. This paper characterises the subset of parameter values, from an a priori set, that satisfy constraints imposed on a directed acyclic graph of constituent functions. The functions and constraints are assumed to be evaluable for given parameter values but need not be known in closed form; they may result from expensive or proprietary simulations, which makes sampling the natural route to an inner approximation of the feasible domain, and dimensionality its principal obstacle. The contribution to artificial intelligence is a class of constraint satisfaction problems defined on directed acyclic graphs, together with forward and backward constraint propagation schemes that solve lower-dimensional subproblems node by node and are proven to yield monotone outer-approximations of the feasible set, and a sampling-based, sampler-agnostic implementation that couples the subproblems through learned surrogate classifiers and regressors, released as open-source software. The engineering application is feasibility analysis of networked systems: the propagated sets tighten the a priori parameter domain before the full problem is solved on the graph. This working hypothesis is demonstrated on four case studies, including a serial batch reactor network, a pharmaceutical tableting line and the verification of a function approximator, where the decomposition reduces the number of constituent function evaluations by up to two orders of magnitude relative to sampling the full problem. Future research will extend the approach to cyclic graphs and account for parametric uncertainty.
\end{abstract}


\end{frontmatter}

\section{Introduction}
\subsection{Background}
Constraint satisfaction problems {(CSP)} are inverse problems that aim to identify a subset of parameter values from an a priori set that is feasible with respect to an assumed constraint system. In general, {CSP} are treated as computational decision problems built on the assumption of mathematical structures that provide an approximate description of an underlying, real-world system. They are an inherent challenge at the core of many disciplines including engineering \citep{song2022learning}, operational research \citep{brailsford1999constraint}, computer science \citep{kumar1992algorithms, ravanbakhsh2015perturbed}, economics and finance \citep{martinez2010systemic}, and the biological and chemical sciences \citep{bodirsky2017complexity}. 

The mathematical structures exploited by a {CSP} explicitly encode the relationship between settings of parameter values within a model, and the constraints imposed on its evaluation either via first principles reasoning or via data-driven modelling. In many cases, the parameters of interest represent decisions to be made within a system's operation; equally, they could represent physical or black-box model parameters.\footnote{Respective examples may include reaction kinetics, and neural network weights and biases.} Once a structure is defined, the model description can be exploited by computational algorithms to identify a set of feasible parameter values that satisfy the constraints imposed. Backtracking search has been shown to be effective for solving {CSP}; however, it aims to return a single parameter value that satisfies constraints imposed on the system \citep{van2006backtracking}. Instead, the following discussion explores algorithms that return solutions to {CSP} with a user-defined minimal cardinality.

{CSPs} often assume evaluation of the model may be described as a composite of constituent functions.\footnote{Many scientific computations are the output of a composite function defined by an evaluation graph or expression tree.} This composition is commonly represented through a directed acyclic graph { (DAG)}, which defines an evaluation order for the constituent functions. However, this {DAG} structure may also be used to represent a network structure within a physical system \citep{jalving2019graph}. For example, much of nature, as well as human constructs, can be thought of and modeled as the interaction between modular entities, each providing a specific function. These entities may themselves have constraints to satisfy. For example in the context of chemical process networks, each unit operation within the network performs a specific physical, chemical or biological transformation \citep{smith2005chemical}. However, the operation itself must satisfy constraints, for example bounds on pressure and temperature that define the specifications of a vessel. These principles may also be applied in analyzing systemic risk within networks of financial entities, such as interbank relationships \citep{boss2004network}, or within the context of quantifying failure probability in structural design problems \citep{zhou2020structural}. 

Many {CSP} provide means to verify the safety of a decision. For example, the field of neural network verification aims to quantify a subset of the domain that a given neural network will provide an accurate approximator to some ground truth function \citep{albarghouthi2021introductionneuralnetworkverification}. However, the solution to a {CSP} is often used thereafter to enhance other decision algorithms. For example, the purpose of constraint propagation within global optimisation solvers is to enhance solution time \citep{puranik2017domain}. The combination of constraint propagation and backtracking inherits the same benefits \citep{van2006backtracking}. 

\subsection{Related Work}
There has been much progress within the development of {CSP} solvers. Within the artificial intelligence community, focus has been on the development of solvers for binary constraint networks defined on variables with finite, discrete domains. For example, the AC3 and AC4 algorithms were developed to obtain arc-consistency \citep{mohr1986arc}. These algorithms are based on enumerating the values within the domains of variables entering into the constraints, but also minimizing computation by pruning values within the variable domains on the fly.

In many real-world problems variables have continuous domains and constraints are defined as functions of many variables. This is a problem setting often approached from the perspective of interval constraint propagation \citep{schichl2005interval}. These methods are based on the use of interval arithmetic to propagate continuous sets through constraint functions and the use of pruning methods to generate an interval outer approximation to the actual feasible set for each variable \citep{benhamou1999revising,paulen2015gpe}.

Examples of {CSP} within the industrial process systems community are provided by the feasibility and flexibility index \citep{swaney1985index_I}, which are applied to assess the operational properties of a candidate process design. This is a concept synonymous with design space characterisation in the pharmaceutical process industry. The approach taken by these works may be summarized as the use of mathematical programming to inscribe a convex region of maximal volume to provide an inner-approximation of the feasible region of operational parameter settings.\footnote{Note that this is desirable here, given the identification of an outer approximation to the feasible region may establish control systems that are inherently unsafe.} However, arriving at such a set requires solving a multi-level mathematical program \citep{ochoa2021novel}. Reformulations may be applied, however solving to global optimality remains a challenge \citep{floudas2001global, mitsos2008global, harwood2017solve}. Additionally, inscription of a simple convex region may lead to significant conservatism in the description of feasibility in practice.

As a result, interest has been directed to the use of adaptive sampling methods. These approaches mitigate the conservatism often present in the approximation of a non-convex region by leveraging the flexibility of sampling to build up a discrete set with a sufficient density of points to elucidate the shape of the feasible region. Examples of adaptive sampling policies are provided by those algorithms based on rejection sampling \citep{kusumo2019bayesian, kucherenko2020computationally}, and search space reduction \citep{sachio2023model}. Parameterisations of the feasible region may in turn be identified by the in-silico data generated by sampling \citep{echard2011ak, boukouvala2012feasibility, kusumo2021design}. A final sampling approach worth noting is that of active learning, where instead one seeks to identify a parameterisation of the feasible region directly; by sampling the space to balance reduction of model uncertainty globally and in regions of interest, such as the boundary of the feasible region \citep{picheny2010adaptive, bect2012sequential}.

The use of sampling lends itself naturally to three {CSP} formulations: nominal, probabilistic and robust. Treatment of systems according to a nominal formulation assumes that all constituent model parameters not involved in the identification are known with certainty. The probabilistic formulation assumes that some parameters not involved in the identification are described according to probability distributions. This then enforces identification of a feasible region that describes chance constraint satisfaction. The application of sampling to the latter is well suited due to ease in use of Monte Carlo approximations to chance constraints. It is worth noting that mathematical programming formulations are also available to handle probabilistic settings, with the caveat that the number of constraints in the underlying model is combinatorial with the number of uncertain parameter scenarios considered \citep{pulsipher2019scalable} and needs to be approached with care. Robust formulations consider identification of designs for the worst-case realizations of uncertain parameters, as demonstrated for the solution of flexibility index problems by \citet{kudva2024robust}. There are equivalences between all formulations under certain specifications of uncertainty and its treatment within the problem definition.

A major barrier to the use of sampling methods for solving {CSP} is the dimensionality of the region one seeks to characterise. The volume of a unit ball grows exponentially with its dimension, which can lead to very low sampling efficiency \citep{buchner2023nested}. This has led to the development of advanced rejection sampling mechanisms \citep{Feroz_2019} that can significantly reduce the total number of model evaluations required to gain a sufficient representation of the feasible region. However, these approaches are in general agnostic to the underlying problem structure, and instead attempt to learn it from model evaluations.
\subsection{Contribution}
We focus on CSPs where knowledge of the function composition is encoded by a network structure. In this case, there is explicit problem structure that may be leveraged to enhance the efficiency of sampling-based approaches to solving a given {CSP}. In the following, we propose new methodology to exploit the network structure that exists in many {CSP} to enhance the sampling efficiency of identification. Our contributions are:
\begin{enumerate}
    \item[{\sf(i)}] Formulation of a class of {CSP}s that are established on a {DAG}.
    \item[{\sf(ii)}] Formalization of constraint propagation techniques that operate with respect to a precedence ordering of the {DAG} descriptive of the {CSP}.
    \item[{\sf(iii)}] Establishment of the contractivity and monotonicity properties of the propagations when applied exactly.
    \item[{\sf(iv)}] Implementation of a sampling-based constraint propagation framework for general constraint functions that is agnostic to the sampling scheme used and available in an open-source Python library.
    \item[{\sf(v)}] Demonstration of the propagations in tightening an a priori parameter domain for sampling solutions to the overall {CSP}, leading to significant improvements in sample efficiency in case study.
\end{enumerate}
The paper is structured as follows: Section \ref{sec:prelim} provides preliminaries on adaptive sampling approaches to {CSP} and constructing surrogates to parameterise feasible sets; Section \ref{sec:method} formulates the problem, then introduces and analyses a constraint propagation strategy leveraging graph decomposition and adaptive sampling; Section \ref{sec:software_imp} details implementation; and Section \ref{sec:CS} demonstrates the method on four case studies, describing {CSP} common in engineering and machine learning. Thorough discussion of the properties of the method, its limitations and the results is provided throughout the paper.

\section{Preliminaries}
\label{sec:prelim}
\subsection{Sampling Approaches to Constraint Satisfaction}
Consider the following problem set-up, where one wishes to identify a subset of parameter values, $\textbf{v} \in \mathbb{V}\subseteq\mathcal{K}_v$ from a compact set $\mathcal{K}_v \subset \mathbb{R}^{n_v}$, that satisfy a system of inequality constraints, 
\begin{equation}\label{eq:DS0}
   \mathbb{V} \coloneqq \{\textbf{v} \in \mathcal{K}_v \mid \mathbf{G}(\textbf{v}) \leq \textbf{0} \},
\end{equation}
with $\mathbf{G}: \mathbb{R}^{n_v} \to \mathbb{R}^{n_g}$. No specific assumption is made on the system of constraints; the evaluation could be algebraic, transcendental or the result of an expensive simulation, such as the solution to a partial differential equation {(PDE)}. Solution to \eqref{eq:DS0} may be identified by solving a model inversion problem, however exact description is generally unavailable even if strong assumptions are made on the set $\mathcal{K}_v$ and the constraint function. 

Sampling approaches provide avenue to approximating \eqref{eq:DS0} and are flexibly applied to general definitions of constraint systems, including those defined by expensive simulators where gradient information may not be readily available. The key idea is to maintain a sampling policy, which generates samples $\textbf{v} \in \mathcal{K}_v$ and a dataset $\mathcal{G}_v = \{(\textbf{v}_k,\textbf{G}(\textbf{v}_k))\}_{k=1}^{K}$ through evaluations made sequentially or in parallel. The policy may either be fixed if it does not utilize information gained from historical function evaluations; or adaptive if it is updated iteratively based on collected data. Policy adaptation is particularly desirable in the spirit of sample efficiency, as underpinned by the schemes developed within the decision theory community. The goals of a sampling policy are ultimately dependent on the decision problem formulated. In the setting considered here, sampling policies are designed to build up a sufficient density of points, 
\begin{equation}\label{eq:sampled_set}
\begin{aligned}
    {\mathbb{V}} \approx \left\{ \textbf{v} \in \mathcal{K}_v \ \middle| \ \exists (\textbf{v}, \textbf{g}) \in \mathcal{G}_v : \ \textbf{g}\leq \textbf{0} \right\},  
\end{aligned}    
\end{equation}
to describe the solution set. One may also become interested in parameterising the set to indicate feasibility in further tasks, and this directs discussion in the following.
\subsection{Learning to Parameterise Feasibility}\label{sec:BC_SM}
One could utilize data generated in sampling $\mathcal{G}_v$ to parameterise a cheap-to-evaluate surrogate function ${\textbf{G}_\varepsilon}$ with some maximum error $\varepsilon > 0$ to the original set parameterisation, such that ${\textbf{G}_\varepsilon} \in {\mathcal{F}_\varepsilon}$ with
\begin{equation*}
    \mathcal{F}_\varepsilon \coloneqq\left\{ {\textbf{G}_\varepsilon}: \mathbb{R}^{n_v} \rightarrow \mathbb{R}^{n_g} \ \middle| \ \forall (\textbf{v}, \textbf{g}) \in \mathcal{G}_v,\ \lvert \textbf{G}_\varepsilon(\textbf{v}) - \textbf{g}\rvert_p \leq \varepsilon \right\},
\end{equation*}
where $\lvert \cdot \rvert_p$ indicates the vector $p$-norm. An indicator function, $\mathbb{I}:\mathbb{R}^{n_g} \rightarrow \{0,1\}$ with $\mathbb{I}[z]=1 \iff z\leq 0$ may then be applied to describe feasibility,
\begin{equation*}
    {{\mathbb{V}}} \approx\left\{ \textbf{v} \in \mathbb{R}^{n_v} \ \middle| \ \mathbb{I}\big[{\textbf{G}_\varepsilon}(\textbf{v})\big] = 1 \right\}.
\end{equation*}
Although this parameterisation is useful for many subsequent tasks, the indicator function is discontinuous, which limits exploitation of gradient-based optimisation solvers. 

Instead, one would prefer binary classifiers to construct an approximation with first-order continuity at least. For instance, a support vector machine (SVM) enables separation of two classes of labeled data, $\mathcal{D}_{s_v} = \{(\textbf{v}_k, s_k)\}_{k=1}^{K}$, where $\textbf{v}\in \mathcal{K}_v $ are datapoints in the domain and the associated labels $s \in \{-1, 1\}$ can be viewed as indications of constraint satisfaction. The possible functions $\overline{\textbf{G}} \in \mathcal{F}$ approximated by this SVM take the form,
\begin{equation*}\label{eq:svm}
    \mathcal{F} = \left\{ \overline{\textbf{G}}:\mathbb{R}^{n_v} \rightarrow \mathbb{R}\ \middle|
    \begin{array}{l}
    \forall (\textbf{v}, s) \in \mathcal{D}_{s_v}: \\ 
    \overline{\textbf{G}}(\textbf{v})\geq 1,\ \text{if $s = 1$} \\
    \overline{\textbf{G}}(\textbf{v}) \leq -1,\ \text{if $s = -1$}
    \end{array}\right\}.
\end{equation*}
The importance of SVM and other binary classifiers in this context is that they provide surrogate approximations of general inequality constraints. Specifically, those points negatively classified by the SVM are deemed to belong to a constraint set that has the same form as \eqref{eq:DS0},
\begin{equation*}
    \mathbb{V} \approx \{ \textbf{v}\in \mathcal{K}_v \mid \overline{\textbf{G}}(\textbf{v}) \leq 0 \}.
\end{equation*}
In practice, the choice of approximator is guided by data availability, among other factors.
\section{Methodology}
\label{sec:method}
\subsection{Constraint satisfaction problems defined on composite functions}
\label{sec:PS}
The formulation of interest is a constraint satisfaction problem ({CSP}) with an assumed structure. Specifically, we assume evaluation of the composite function may be described as a directed acyclic graph ({DAG}), $\mathcal{G} = (\mathcal{N}, \mathcal{E})$, where $\mathcal{N} = \{1,\ldots, N\}$ is an ordered set of nodes and $(j,i) \in \mathcal{E}$ is the set of directed edges, such that $j \prec i$ indicates a partial ordering of the nodes.\footnote{Identifying an ordering of the nodes is straightforward for small acyclic graphs. For larger {DAGs}, this task may be automated, e.g. using the Coffman-Graham algorithm, which is commonly applied to job-shop scheduling problems under precedence constraints between tasks \citep{coffman1972optimal}.} The connectivity of the {DAG} may be described equivalently by a non-symmetric adjacency matrix $A\in \{0,1\}^{N\times N}$, with element $a_{ik} =1$ indicating a connection from node $i$ to node $k$.
The following data are assumed for each node $i\in \mathcal{N}$: 
\begin{itemize}
    \item In-neighbours, $j \in \mathcal{N}_i^{\rm in}$, which provide parent nodes to node $i$, given by $\mathcal{N}^{\rm in}_i \coloneqq \{ j \mid (j,i) \in \mathcal{E}, j \prec i \} $;
    \item Out-neighbours, $k \in \mathcal{N}^{\rm out}_i$, for which node \textit{i} provides a parent node, given by $\mathcal{N}^{\rm out}_i \coloneqq \{ k \mid (i,k) \in \mathcal{E}, i\prec k\} $;   
    \item Input variables to node $i$ from any parent $j \in \mathcal{N}_i^{\rm in}$, denoted by $\textbf{u}^j_i \in \mathbb{R}^{n_{u_{ij}}}$;
    \item Output variables from node $i$ to any descendant $k\in \mathcal{N}^{\rm out}_i$, denoted by $\textbf{y}_i^k \in \mathbb{R}^{n_{y_{ik}}}$;
    \item Constituent relations between the output variables of node $i$ connected to a descendant $k\in \mathcal{N}^{\rm out}_i$ and its input variables,
    \begin{equation}\label{eq:Fik}
        \textbf{y}_i^k = \textbf{F}_i^k(\textbf{v}_i, \textbf{u}_i)\,,
    \end{equation}
    where $\textbf{v}_i \in \mathbb{R}^{n_{v_i}}$ are parameters and $\textbf{u}_i \coloneqq (\textbf{u}_i^j)_{j \in \mathcal{N}^{\rm in}_i} \in \mathbb{R}^{n_{u_i}}$ denotes the concatenation of all input variables to node $i$.\footnote{The mappings $\textbf{F}_i^k$ are assumed to be deterministic but need not be given in closed-form. Where convenient, $\textbf{y}_i=\textbf{F}_i(\textbf{v}_i, \textbf{u}_i)$ will be denoted as a single vector-valued function that describes the outlet variables $\textbf{y}_i$.} It is important to note that $\textbf{v}_i$ are assumed local to node ${i}\in \mathcal{N}$. Section \ref{sec:extensions} details an extension to CSP where additional coupling parameters appear within the formulation.
    \item Any inequality constraints to be satisfied by the local parameters and input variables to node $i$,
    \begin{equation}\label{eq:Gi}
        \textbf{G}_i(\textbf{v}_i, \textbf{u}_i) \coloneqq \textbf{g}_i(\textbf{v}_i, \textbf{u}_i, \textbf{y}_i) \leq \textbf{0}\,,
    \end{equation}
    where $\textbf{y}_i \coloneqq (\textbf{y}^k_i)_{k\in\mathcal{N}_i^{\rm out}}\in \mathbb{R}^{n_{y_i}}$ denotes the concatenation of all output variables from node \textit{i}.
\end{itemize}
The challenge that is posed by this CSP is to identify the values of $\textbf{v} \coloneq (\textbf{v}_i)_{i\in \mathcal{N}} \in \mathbb{R}^{n_{v}}$ that satisfy the constraints imposed on all of the local node variables. Formally, such a set may be defined as
\begin{equation}\label{eq:DS_joint}
    \mathbb{V} \coloneqq \left\{ \textbf{v} \in \mathcal{K}_{v}\ \middle|
    \begin{array}{l}
        \forall i\in \mathcal{N}, \exists (\mathbf{u}_i,\mathbf{y}_i)\in \mathbb{R}^{n_{u_i}}\times\mathbb{R}^{n_{y_i}}:\\
        \mathbf{G}_i(\textbf{v}_i, \textbf{u}_i) \leq \textbf{0}\\
        \textbf{y}_i = \textbf{F}_i(\textbf{v}_i, \textbf{u}_i),\\
        \textbf{u}_k^i = \textbf{y}_i^k,\ \forall k\in\mathcal{N}^{\rm out}_i\\
    \end{array} \right\},
\end{equation}
where $\mathcal{K}_{v} \coloneqq \prod_{i\in\mathcal{N}} \mathcal{K}_{v_i} \subset \mathbb{R}^{n_v}$ describes the search domain of local node parameters $\mathbf{v}_i \in \mathcal{K}_{v_i}$, and reflects a priori knowledge about the problem at hand. The only assumption is that the constraint evaluation is defined to respect the precedence ordering established on evaluation of the composite function itself. Under the assumption that the node input variables $\textbf{u}_i$, $i \in \mathcal{N}$ are set by selection of the values of parameters $\textbf{v}$, the right-hand side of \eqref{eq:DS_joint} could be rewritten as in \eqref{eq:DS0}. The following example is provided to help guide the reader through presentation of the methodology.

\begin{figure}[htb]
    \centering
    \includegraphics[width=\linewidth]{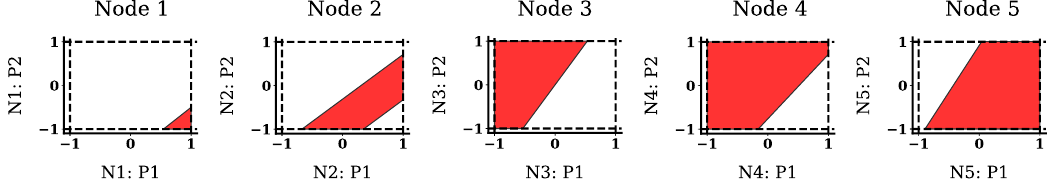}
    \caption{Projections of the solution set \eqref{eq:DS_joint} (red) for Example \ref{ex:1} onto the subspace of node parameters (1--5, left--right) identified through polyhedral operations implemented in {\sf MPT 4.0}. The subplot axes labels reference node parameters with $\sf Ni: Pj$ denoting the \textit{j}\textsuperscript{th} decision parameter at \textit{i}\textsuperscript{th} node, $v_{ij}$.}
    \label{fig:ex1projections}
\end{figure}

\begin{example}\label{ex:1}
    Consider the following {DAG} of a composite function
\begin{center}
    \begin{tikzpicture}
    \node[draw, circle] (1) at (0,-0.5) {1};
    \node[draw, circle] (2) at (0,0.5) {2};
    \node[draw, circle] (3) at (2,0) {3};
    \node[draw, circle] (4) at (4,0.5) {4};
    \node[draw, circle] (5) at (4,-0.5) {5};
    \draw[->] (1) -- (3);
    \draw[->] (2) -- (3);
    \draw[->] (3) -- (4);
    \draw[->] (3) -- (5);
    \end{tikzpicture}
\end{center}
where $\mathcal{N}_1^{\rm in} = \mathcal{N}_2^{\rm in} = \emptyset$, $\mathcal{N}_1^{\rm out}= \mathcal{N}_2^{\rm out}= \{3\}$, $\mathcal{N}_3^{\rm in} = \{1,2\}$, $\mathcal{N}_3^{\rm out} = \{4,5\}$, $\mathcal{N}_4^{\rm in} = \mathcal{N}_5^{\rm in} =\{3\}$, and $\mathcal{N}_4^{\rm out} = \mathcal{N}_5^{\rm out} =\emptyset$, and with the precedence ordering $1\prec 2 \prec 3 \prec 4 \prec 5$. Each node is associated with an affine function relating its inputs ${\bf u}_i \in \mathbb{R}^{\lvert\mathcal{N}_i^{\rm in}\rvert}$ and parameters ${\bf v}_i\in \mathbb{R}^2$ to the outputs ${y}^k_i\in \mathbb{R}$ directed to node $k\in \mathcal{N}_i^{\rm out}$,
\begin{equation}\label{eq:ex1_forward}
    {y}^k_i = {F}^k_i({\bf v}_i, {\bf u}_i) \coloneqq A^k_i{\bf v}_i + B^k_i{\bf u}_i,
\end{equation}
with $A^k_i \in \mathbb{R}^{1 \times 2}$ and $B^k_i\in \mathbb{R}^{1 \times \lvert\mathcal{N}_i^{\rm in}\rvert}$. Each node $\textit{i} \in \mathcal{N}$ has constraint functions imposed on output variables, 
\begin{equation}\label{eq:ex1_constraint}
    G_i({\bf v}_i, {\bf u}_i) = F_i({\bf v}_i, {\bf u}_i) \leq {\bf 0} \,.
\end{equation}
The search domains for each node are defined by the boxes $\mathcal{K}_{v_i}=[-1,1]^2$. Due to the linearity of the constituent functions~\eqref{eq:ex1_forward} and constraints~\eqref{eq:ex1_constraint}, the set $\mathbb{V}$ in \eqref{eq:DS_joint} for this example is available analytically. For instance, numerical solution for the projections of $\mathbb{V}$ onto the subspace of node-wise parameters is available through polyhedral operations using the Multiparametric Toolbox {\sf MPT 4.0} in {\sf MATLAB} \citep{mpt}. The projections are detailed by Figure~\ref{fig:ex1projections}. 
\end{example}

In the case of nonlinear functions in \eqref{eq:Fik} or \eqref{eq:Gi}, an exact characterisation of $\mathbb{V}$ is no longer possible in general, and a sampling algorithm needs to be applied to compute an inner-approximation to $\mathbb{V}$ instead. We refer to this as the simultaneous approach. Clearly, the combined dimensionality $n_{v} = \sum_{i\in\mathcal{N}} n_{v_i}$ of $\mathcal{K}_v$ must be sufficiently small for problem \eqref{eq:DS_joint} to remain directly tractable using sampling.

To address these scalability challenges, we propose to leverage the composite structure available to effectively reduce the a priori search domain $\mathcal{K}_v$, before attempting the simultaneous estimation of $\mathbb{V}$.

\subsection{Decomposition Approach}
\label{sec:decomp}

We explore methodology to decompose the parameter estimation into separate subproblems, to be solved incrementally. The aim is for this incremental identification to determine a tight enclosure of the feasible set $\mathbb{V}$ in \eqref{eq:DS_joint}, which may be used as a reduced domain for subsequent (simultaneous) parameter estimation. The working hypothesis is that decomposing the problem into a series of subproblems with lower dimensionality will improve the overall efficiency of the parameter estimation task. 

The methodology is detailed in the following subsections. To aid with explanation, we continue the illustrative Example \ref{ex:1}. All code can be found in the associated code base $\boldsymbol{\mu}\textbf{F}$.\footnote{\ifanon Repository URL withheld for review.\else\url{https://github.com/mawbray/mu.F}\fi}

\subsubsection{Local Subproblems}
The problem structure described in Section~\ref{sec:PS} admits a node-wise decomposition because the inequality constraints \eqref{eq:Gi} are imposed on each constituent function in terms of local node variables only. In particular, due to the coupling between nodes this leads to a block diagonal structure in the biadjacency matrix of a corresponding variable-constraint bipartite network \citep{del2016automated, daoutidis2019decomposition} representing the simultaneous problem as defined in \eqref{eq:DS_joint}.

We propose a decomposition methodology to identify an approximation to the local feasible set $\mathbb{VU}_i$ of node $i\in \mathcal{N}$, given by
\begin{equation}\label{eq:DS_unit2}
    \mathbb{VU}_i \coloneqq \left\{ (\textbf{v}_i,\textbf{u}_i) \in \mathcal{K}_{vu_i}\ \middle|
    \begin{array}{l}
        \mathbf{0} \geq \mathbf{G}_i(\textbf{v}_i, \textbf{u}_i)\\[2mm]
        \forall k\in\mathcal{N}^{\rm out}_i,\ \exists (\textbf{v}_k,\textbf{u}_k) \in \mathbb{VU}_k:\\
        \quad \textbf{u}_k^i = \textbf{F}_i^k(\textbf{v}_i, \textbf{u}_i)\\[2mm]
        \forall j\in\mathcal{N}^{\rm in}_i,\ \exists (\textbf{v}_j,\textbf{u}_j) \in \mathbb{VU}_j:\\ 
        \quad \textbf{u}_i^j = \textbf{F}_j^i(\textbf{v}_j, \textbf{u}_j)\\
    \end{array} \right\}\,,
\end{equation}
which describes the feasible local parameters $\mathbf{v}_i$ and input variables $\mathbf{u}_i$ within the set $\mathcal{K}_{vu_i}\coloneqq \mathcal{K}_{v_i} \times \mathcal{K}_{u_i}$. From the definition \eqref{eq:DS_unit2}, the local feasible set satisfies constraints imposed on the node locally $\textbf{G}_i(\textbf{v}_i, \textbf{u}_i) \leq \textbf{0}$, as well as those inherited from the parent and descendant nodes. By construction, the projection of $\mathbb{VU}_i$ onto local node parameters $\textbf{v}_i$, given by
\begin{equation}
    \mathbb{V}_i = \left\{\textbf{v}_i \in \mathcal{K}_{v_i}~\middle| \begin{array}{l}
           \exists \textbf{u}_i \in \mathcal{K}_{u_i}: (\textbf{v}_i, \textbf{u}_i) \in \mathbb{VU}_i
    \end{array}\right\}\,,
\end{equation}
contains the projection of $\mathbb{V}$ onto $\textbf{v}_i$ denoted $\text{proj}_{\textbf{v}_i} \mathbb{V}$, and we have $\mathbb{V} \subseteq \prod_{i\in \mathcal{N}} \mathbb{V}_i$. It is worth noting that {(\sf i)} no statement on the converse inclusion is provided and {(\sf ii)} that the very incorporation of the input variables $\textbf{u}_i$ into the local node problem provides the base for problem decomposition.

\subsubsection{Solution Methodology}\label{sec:overview_dec}

Observe that solution to the subproblem \eqref{eq:DS_unit2} defined on a node ${i} \in \mathcal{N}$ requires knowledge of the solutions to all subproblems. For a {DAG} possessing any node with non-zero degree (i.e. with any directed connection to another node), these subproblems cannot be solved directly as a result. Nevertheless, we can obtain a relaxation (outer-approximation) to \eqref{eq:DS_unit2} by deleting selected constraints. Two options for defining a tractable relaxation entail deleting constraints coupling a node either to its out-neighbours or to its in-neighbours. We term these approximations the {\em forward relaxation} and the {\em backward relaxation}, respectively. Both approximations to the subproblems can be solved incrementally, according to the precedence ordering on the nodes in the case of the forward relaxation and to the reverse of the precedence ordering in the case of the backward relaxation. 

\paragraph{Forward Relaxation and Propagation}
Subproblems may be relaxed by deleting constraints that couple a node $i\in \mathcal{N}$ to its out-neighbours $k \in \mathcal{N}_i^{\rm \text{out}}$ to provide the forward relaxation ${\mathbb{VU}}^{({\sf f})}_i \supseteq \mathbb{VU}_i$, given by
\begin{equation}\label{eq:DS_unitFP0}
    {\mathbb{VU}}^{({\sf f})}_i \coloneqq \left\{ (\textbf{v}_i,\textbf{u}_i) \in \mathcal{K}^{({\sf f})}_{vu_i}\ \middle|
    \begin{array}{l}
        \mathbf{0} \geq \mathbf{G}_i(\textbf{v}_i, \textbf{u}_i)\\[2mm]
        \forall j\in\mathcal{N}^{\rm in}_i,\ \exists (\textbf{v}_j,\textbf{u}_j) \in {\mathbb{VU}}^{({\sf f})}_j:\\
        \quad {\sf (a)} \ \textbf{u}_i^j = \textbf{F}_j^i(\textbf{v}_j, \textbf{u}_j)\\
        \quad {\sf (b)} \ \forall k \in \mathcal{N}_j^{\rm out}, k \prec i,\ \exists (\textbf{v}_k, \textbf{u}_k) \in \mathbb{VU}_k^{({\sf f})}: \\
        \quad \quad \quad \textbf{u}_k^j = \textbf{F}_j^k(\textbf{v}_j, \textbf{u}_j)
    \end{array} \right\}.
\end{equation}
The search domain $\mathcal{K}^{({\sf f})}_{vu_i} \coloneqq \mathcal{K}_{v_i} \times \mathcal{K}_{u_i}^{(\sf f)}$ is taken as the Cartesian product of the images of ${\mathbb{VU}}^{(\sf f)}_j$ under the forward evaluation of the constituent functions of in-neighbours $j \in \mathcal{N}_i^{\sf \text{in}}$, such that $\mathcal{K}_{u_i}^{(\sf f)}=\prod_{j\in\mathcal{N}_i^{\sf \text{in}}}\textbf{F}_j^i[\mathbb{VU}^{(\sf f)}_j]$.  

The sets ${\mathbb{VU}}_i^{(\sf f)}$ are then identified by solving subproblems according to the precedence ordering of the {DAG} and propagating the result forward for the subsequent subproblem solution. To provide intuition as to why this is the case, note those nodes that have zero in-degree, also called root nodes $i \in \mathcal{N}^{\rm R}$, can be identified independently from other nodes,
\begin{equation*}\label{eq:VUroot}
    {\mathbb{VU}}^{({\sf f})}_{i} \coloneqq \{\textbf{v}_{i} \in \mathcal{K}^{({\sf f})}_{{v}_{i}} \mid \mathbf{0} \geq \textbf{G}
_{i}(\textbf{v}_{i}, \textbf{u}_i) \}, \ \forall i\in \mathcal{N}^{\rm R}.
\end{equation*}
Additionally, a given pair $(\textbf{v}_i, \textbf{u}_i) \in {\mathbb{VU}}^{\sf (f)}_i$ ensures: {\sf (i)} the local node constraints, $\textbf{G}_i(\textbf{v}_i, \textbf{u}_i)\leq \textbf{0}$ are satisfied, and {\sf (ii)} the input variables $\textbf{u}_i$ to node \textit{i} are defined by feasible selection of $(\textbf{v}_j, \textbf{u}_j) \in {\mathbb{VU}}^{\sf (f)}_j$,  $\forall j \prec i$. The coupling constraints ({\sf a}) and ({\sf b}) in \eqref{eq:DS_unitFP0} enforce restrictions on the input variables to node \textit{i} according to the forward relaxations of node $j\in \mathcal{N}_i^{\rm in}$ and those nodes $k\in \mathcal{N}_j^{\rm out}$, $k\prec i$ with input from node $j$, respectively. In particular, the constraint ({\sf b}) provides additional tightening when the graph has structure such that $\lvert\mathcal{N}_j^{\rm out}\rvert > 1$ for some $j \in \mathcal{N}_i^{\rm in}$.

\begin{continuedexample}{ex:1}
The forward propagation and projections of the solution set onto the subspace of node-wise parameters are shown in Figure \ref{fig:approx1_proj}. It is clear from comparison to Figure \ref{fig:ex1projections} that the projections are indeed supersets to the projections of the exact solution, most noticeably for node 2. 
\begin{figure}[hb!]
    \centering
    \includegraphics[width=\linewidth]{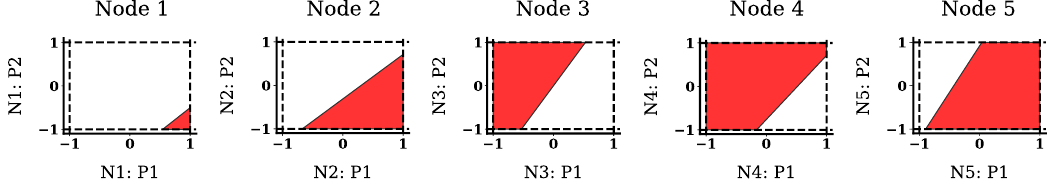}
    \caption{Projections of the forward propagation solution \eqref{eq:DS_unitFP0} (red) for Example \ref{ex:1} onto the subspace of node parameters (1--5, left--right) identified through polyhedral operations implemented in {\sf MPT 4.0}.}
    \label{fig:approx1_proj}
\end{figure}
\end{continuedexample}

\paragraph{Backward Relaxation and Propagation}
Subproblems may similarly be relaxed by deleting the constraints coupling a node \textit{i} to its in-neighbours $j \in \mathcal{N}_i^{\rm in}$ and then employing a policy of backpropagation,
\begin{equation}\label{eq:DS_unitBP}
    {\mathbb{VU}}^{({\sf b})}_i \coloneqq \left\{ (\textbf{v}_i,\textbf{u}_i) \in \mathcal{K}^{({\sf b})}_{vu_i}\ \middle|
    \begin{array}{l}
        \mathbf{0} \geq \mathbf{G}_i(\textbf{v}_i, \textbf{u}_i)\\[2mm] 
        \forall k\in\mathcal{N}^{\rm out}_i,\ \exists (\textbf{v}_k,\textbf{u}_k) \in {\mathbb{VU}}^{({\sf b})}_k:\\
        \quad ({\sf a}) \ \textbf{u}_k^i = \textbf{F}_i^k(\textbf{v}_i, \textbf{u}_i)\\
        \quad ({\sf b}) \ \forall j \in \mathcal{N}_k^{\rm in}, j \succ i,\ \exists (\textbf{v}_j, \textbf{u}_j) \in \mathbb{VU}_j^{\sf (b)}:\\ 
        \quad \quad \quad \textbf{u}_k^j = \textbf{F}_j^k(\textbf{v}_j, \textbf{u}_j)
    \end{array} \right\},
\end{equation}
with $\mathcal{K}^{({\sf b})}_{{vu}_i} \coloneqq \mathcal{K}_{{v}_i} \times \mathcal{K}^{({\sf b})}_{{u}_i}$ and ${\mathbb{VU}}^{({\sf b})}_i \supseteq {\mathbb{VU}}_i$. The set $\mathcal{K}^{({\sf b})}_{{u}_i}$ indicates the search domain of local input variables to node $i$, and can be estimated as the Cartesian product of the images of $\mathcal{K}^{\sf (b)}_{{vu}_j}$ under the forward evaluation of the constituent functions of node $j \in \mathcal{N}_i^{\rm in}$, $\mathcal{K}_{u_i}^{\sf (b)} = \prod_{j\in \mathcal{N}_i^{\rm in}} \textbf{F}_j^i[\mathcal{K}^{\sf (b)}_{vu_j}]$. The sets $\mathcal{K}^{\sf (b)}_{u_j}, \forall j \in \mathcal{N}_i^{\rm in}$ may be estimated from their in-neighbours and so on until a root node is reached. Practically, this estimation process proceeds forward according to the precedence ordering of the nodes, prior to application of the backward relaxation and propagation.\footnote{It is worth noting that unlike the search domain $\mathcal{K}^{\sf (f)}_{u_i
}$ in the forward relaxation \eqref{eq:DS_unitFP0}, the a priori set $\mathcal{K}^{\sf (b)}_{u_i
}$ in \eqref{eq:DS_unitBP} could include input variables reachable from infeasible settings of $\textbf{v}_j \in \mathcal{K}_{v_j} \ \forall j \prec i$. Therefore, one expects $\mathcal{K}^{\sf (f)}_{u_i
} \subseteq \mathcal{K}^{\sf (b)}_{u_i
}$ when a backward relaxation and propagation is first applied.}

The intuition for the use of backwards propagation is that the leaf nodes in the {DAG}, $i \in \mathcal{N}^{\rm L}$ are the only nodes from which one can initially solve a subproblem~\eqref{eq:DS_unitBP}. This is because they possess no out-neighbors and, therefore, no complicating constraints,
\begin{equation*}\label{eq:VUleaf}
    {\mathbb{VU}}^{({\sf b})}_{i} \coloneqq \{(\textbf{v}_{i}, \textbf{u}_{i}) \in \mathcal{K}^{({\sf b})}_{{vu}_{i}} \mid \mathbf{0} \geq \textbf{G}
_{i}(\textbf{v}_{i}, \textbf{u}_{i}) \}, \ \forall i \in \mathcal{N}^{\rm L} \,.
\end{equation*}
Then, the backward relaxation ensures that for any pair $(\textbf{v}_i, \textbf{u}_i) \in {\mathbb{VU}}^{({\sf b})}_i$: ({\sf i}) the local constraints $\textbf{G}_i(\textbf{v}_i, \textbf{u}_i) \leq \textbf{0}$ are satisfied, and ({\sf ii}) for all output nodes $k \in \mathcal{N}_i^{\text{out}}$ there is a setting $(\textbf{v}_k, \textbf{u}_k)$ under which the local unit constraints $\textbf{G}_k(\textbf{v}_k, \textbf{u}_k) \leq \textbf{0}$ are satisfied. The coupling constraints ({\sf a}) and ({\sf b}) in \eqref{eq:DS_unitBP} enforce restrictions on the node \textit{i} output variables according to the backward relaxation of node $k\in \mathcal{N}_i^{\rm out}$ and those nodes $j \in \mathcal{N}_k^{\rm in}$ providing input to node $k$ with ordering after node $i\prec j$, respectively. This time the second constraint ({\sf b}) provides additional tightening when the graph has structure such that  $\lvert\mathcal{N}_i^{\rm in}\rvert>1$ for some $k \in \mathcal{N}_i^{\rm out}$.

\begin{continuedexample}{ex:1}
    The backward propagation and projections of the solution set onto the subspace of node-wise parameters for Example \ref{ex:1} are shown in Figure \ref{fig:approx2_proj}. It is clear from comparison to Figure \ref{fig:ex1projections} that the projections are supersets to the projections of the exact solution, most noticeably for nodes 4 and 5. 

\begin{figure}[ht!]
    \centering
    \includegraphics[width=\linewidth]{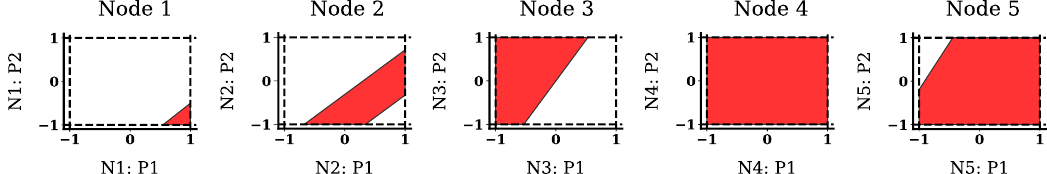}
    \caption{Projections of the backward propagation solution \eqref{eq:DS_unitBP} (red) for Example \ref{ex:1} onto the subspace of node parameters (1--5, left--right) identified through polyhedral operations implemented in {\sf MPT 4.0}.}
    \label{fig:approx2_proj}
\end{figure}
\end{continuedexample}

\paragraph{Composing Propagations}

Relaxing \eqref{eq:DS_unit2} and solving according to \eqref{eq:DS_unitFP0} or \eqref{eq:DS_unitBP} provides an initial outer-approximation. This outer-approximation may then be further tightened by iterating in the opposing direction to the initial propagation with respect to the precedence ordering. In practice, it is convenient to first proceed forwards and then backwards, since the a priori sets $\mathcal{K}_{u_i}^{\sf (f)}$ are initialized more easily than their counterparts $\mathcal{K}_{u_i}^{\sf(b)}$. Then, the backward propagation becomes
\begin{equation}\label{eq:DS_unitFBP}
    {\mathbb{VU}}^{({\sf fb})}_i \coloneqq \left\{ (\textbf{v}_i,\textbf{u}_i) \in \mathbb{VU}^{({\sf f})}_{i} \middle|
    \begin{array}{l}
        \mathbf{0} \geq \mathbf{G}_i(\textbf{v}_i, \textbf{u}_i)\\[2mm] 
        \forall j\in\mathcal{N}^{\rm in}_i,\ \exists (\textbf{v}_j,\textbf{u}_j) \in {\mathbb{VU}}^{({\sf f})}_j:\\
        \quad {\sf (a)} \ \textbf{u}_i^j = \textbf{F}_j^i(\textbf{v}_j, \textbf{u}_j)\\
        \quad {\sf (b)} \ \forall k \in \mathcal{N}_j^{\rm out}, l \prec i,\ \exists (\textbf{v}_k, \textbf{u}_k) \in \mathbb{VU}_k^{({\sf f})}: \\
        \quad \quad \quad \textbf{u}_k^j = \textbf{F}_j^k(\textbf{v}_j, \textbf{u}_j)\\[2mm]
        \forall k\in\mathcal{N}^{\rm out}_i,\ \exists (\textbf{v}_k,\textbf{u}_k) \in {\mathbb{VU}}^{({\sf fb})}_k:\\
        \quad (\sf a') \ \textbf{u}_k^i = \textbf{F}_i^k(\textbf{v}_i, \textbf{u}_i)\\
        \quad ({\sf b'}) \ \forall j \in \mathcal{N}_k^{\rm in}, j \succ i,\ \exists (\textbf{v}_j, \textbf{u}_j) \in \mathbb{VU}_j^{\sf (fb)}:\\ 
        \quad \quad \quad \textbf{u}_k^j = \textbf{F}_j^k(\textbf{v}_j, \textbf{u}_j)
    \end{array} \right\}.
\end{equation}
Since the backward propagation \eqref{eq:DS_unitFBP} enforces additional constraints compared to \eqref{eq:DS_unitFP0}, it follows that ${\mathbb{VU}}^{(\sf fb)}_i\subseteq {\mathbb{VU}}_i^{(\sf f)}$. 

It is worth highlighting that an equivalent problem on ${\mathbb{VU}}^{(\sf bf)}_i$ entails the application of a backward propagation first and a forward propagation second. The composition of the second propagation could even be applied to a subgraph without the need to start from a leaf. One could also define arbitrary compositions of propagations, with a forward-backward-forward iteration, for instance, yielding the set ${\mathbb{VU}}^{({\sf fbf})}_i$. Further discussion of these cases is omitted for conciseness. \bigskip

\begin{continuedexample}{ex:1}
    The forward-backward propagation solution is illustrated in Figure~\ref{fig:approxn_proj}. Comparison to both Figures \ref{fig:approx1_proj} and \ref{fig:approx2_proj} indicates that an additional reduction has been made to the previous propagations. Comparison to Figure \ref{fig:ex1projections} shows that, for this example, the projections after the combined pass coincide with the projections of $\mathbb{V}$.

\begin{figure}[ht!]
    \centering
    \includegraphics[width=\linewidth]{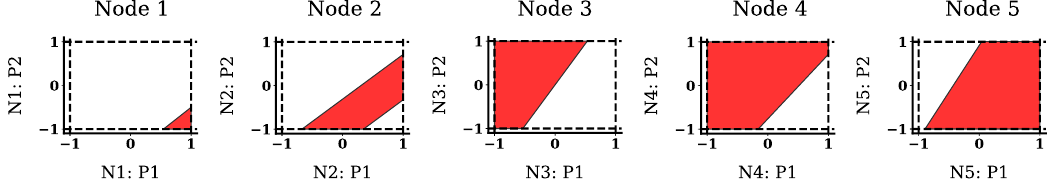}
    \caption{Projections of the forward-backward propagation solution \eqref{eq:DS_unitFBP} (red) for Example \ref{ex:1} onto the subspace of node parameters (1--5, left--right) identified through polyhedral operations implemented in {\sf MPT 4.0}.}
    \label{fig:approxn_proj}
\end{figure}
\end{continuedexample}

\paragraph{Inclusion and Convergence Properties}
Under the following assumptions:
\begin{enumerate}
    \item[{\sf (i)}] the search domain $\mathcal{K}_v\subset\mathbb{R}^{n_{v}}$ is a compact set,
    \item[{\sf (ii)}] both the constituent functions $\textbf{F}_i^k$ and constraint functions $\textbf{G}_i$ are continuous,
\end{enumerate}
the input domains $\mathcal{K}_{u_i}$, combined search domains $\mathcal{K}_{{vu}_i}$, and local feasible sets ${\mathbb{VU}}_i$ in \eqref{eq:DS_unit2} are all compact. 

Since the forward relaxation ${\mathbb{VU}}^{({\sf f})}_i$ in \eqref{eq:DS_unitFP0} proceeds by deletion of the out-neighbour constraints from ${\mathbb{VU}}_i$ in \eqref{eq:DS_unit2}, and under the assumption that $\mathcal{K}^{({\sf f})}_{{vu}_i}\supseteq \mathcal{K}_{{vu}_i}$ and compact, a finite induction according to the precedence ordering of nodes $i\in\mathcal{N}$ shows that ${\mathbb{VU}}^{({\sf f})}_i\supseteq {\mathbb{VU}}_i$ and compact. Consequently, the projections of these sets onto the subspace of local node parameters are also compact and,
\begin{equation*}
    {\mathbb{V}}^{({\sf f})}_i \coloneqq \left\{ \textbf{v}_i \in \mathcal{K}_{v_i}\ \middle|
    \begin{array}{l}
        \exists \textbf{u}_i \in \mathcal{K}^{(\sf f)}_{u_i}:
        (\textbf{v}_i,\textbf{u}_i) \in {\mathbb{VU}}^{({\sf f})}_i
    \end{array} \right\} \supseteq \mathbb{V}_i.
\end{equation*}
Similarly, a finite induction according to the reverse precedence ordering concludes that ${\mathbb{VU}}^{({\sf b})}_i\supseteq {\mathbb{VU}}_i$ whenever $\mathcal{K}^{({\sf b})}_{{vu}_i}\supseteq \mathcal{K}_{{vu}_i}$ for each $i\in\mathcal{N}$, and thus, 
\begin{equation*}
    {\mathbb{V}}^{({\sf b})}_i \coloneqq \left\{ \textbf{v}_i \in \mathcal{K}_{v_i}\ \middle|
    \begin{array}{l}
        \exists \textbf{u}_i \in \mathcal{K}^{(\sf b)}_{u_i}:
        (\textbf{v}_i,\textbf{u}_i) \in {\mathbb{VU}}^{({\sf b})}_i
    \end{array} \right\} \supseteq \mathbb{V}_i,
\end{equation*}
with all these sets being compact. Next, in composing the relaxations and propagations as for ${\mathbb{VU}}^{({\sf fb})}_i$ in \eqref{eq:DS_unitFBP}, we retain the constraints from the previous propagation that describe an outer-approximation of ${\mathbb{VU}}_i$, while adding new constraints that also describe an outer-approximation of ${\mathbb{VU}}_i$ or tightening the domains of participating parameters and input variables. Therefore, the resulting local feasible sets are themselves guaranteed outer-approximations, ${\mathbb{VU}}^{({\sf fb})}_i\supseteq {\mathbb{VU}}_i$, and simultaneously tighten the local feasible sets from previous propagations, ${\mathbb{VU}}^{({\sf fb})}_i\subseteq {\mathbb{VU}}^{({\sf f})}_i$.

The following property summarises the inclusion sequence obtained through alternating composition of forward and backward propagations.
\begin{property}\label{prop:monotone_inclusion}
If $\mathcal{K}^{({\sf f})}_{{vu}_i}\supseteq \mathcal{K}_{{vu}_i}$, the relaxation and propagation of local feasible sets and their projections onto the subspace of local node parameters define monotone inclusion sequences,
\begin{align*}
    & \mathcal{K}^{({\sf f})}_{{vu}_i}\supseteq {\mathbb{VU}}^{({\sf f})}_i\supseteq {\mathbb{VU}}^{({\sf fb})}_i\supseteq {\mathbb{VU}}^{({\sf fbf})}_i\supseteq \cdots \supseteq {\mathbb{VU}}_i\\
    & \mathcal{K}_{{v}_i}\supseteq {\mathbb{V}}^{({\sf f})}_i\supseteq {\mathbb{V}}^{({\sf fb})}_i\supseteq {\mathbb{V}}^{({\sf fbf})}_i\supseteq \cdots \supseteq {\mathbb{V}}_i \supseteq \text{proj}_{\textbf{v}_i} \mathbb{V}.
\end{align*}
Likewise, if $\mathcal{K}^{({\sf b})}_{{vu}_i}\supseteq \mathcal{K}_{{vu}_i}$, the relaxation and propagation of local feasible sets and their projections onto the subspace of local node parameters define monotone inclusion sequences,
\begin{align*}
    & \mathcal{K}^{({\sf b})}_{{vu}_i}\supseteq {\mathbb{VU}}^{({\sf b})}_i\supseteq {\mathbb{VU}}^{({\sf bf})}_i\supseteq {\mathbb{VU}}^{({\sf bfb})}_i\supseteq \cdots \supseteq {\mathbb{VU}}_i\\
    & \mathcal{K}_{{v}_i}\supseteq {\mathbb{V}}^{({\sf b})}_i\supseteq {\mathbb{V}}^{({\sf bf})}_i\supseteq {\mathbb{V}}^{({\sf bfb})}_i\supseteq \cdots \supseteq {\mathbb{V}}_i \supseteq \text{proj}_{\textbf{v}_i} \mathbb{V} .
\end{align*}
\end{property}

A direct consequence of the nonincreasing set sequences in Property~\ref{prop:monotone_inclusion} is the existence of the set limits $\lim_{d\to\infty} {\mathbb{VU}}^{({\sf d})}_i \supseteq {\mathbb{VU}}_i$ and $\lim_{d\to\infty} {\mathbb{V}}^{({\sf d})}_i \supseteq {\mathbb{V}}_i$ for both ${\sf d}\in\{{\sf f}, {\sf fb}, {\sf fbf}, \ldots\}$ or ${\sf d}\in\{{\sf b}, {\sf bf}, {\sf bfb}, \ldots\}$. Yet, Property \ref{prop:monotone_inclusion} alone does not guarantee for the corresponding set limits to be ${\mathbb{VU}}_i$ or $ {\mathbb{V}}_i$ or establish a rate for their convergence, which will likely call for stronger assumptions on the problem setting \citep{aubin1990set, rockafellar1998set}. 

In Example \ref{ex:1}, the forward-backward and backward-forward
compositions were observed to yield $\mathbb{V}^{\sf (fb)}_i = \mathbb{V}^{\sf (bf)}_i$ for all $i \in \mathcal{N}$ after a single pass in each
direction. This behavior should not be expected in general: the set limits remain outer-approximations of
$\mathbb{V}_i$ and of the projections of $\mathbb{V}$. The practical value of the compositions lies in the monotone
tightening of the search domain established by Property~\ref{prop:monotone_inclusion}.

\subsubsection{Methodology Implementation}\label{sec:general_sol}

The previous section presented a general methodology to approximate $\mathbb{VU}_i$ but omitted practical detail of how to compute these propagations. In this section, focus is directed towards a practical implementation of these propagations in various problem settings. In particular, our implementation addresses three key challenges:
\begin{enumerate}
    \item[{\sf (i)}] identifying a parameterisation of the local feasible set ${\mathbb{VU}}_i^{(\sf d)}$ in the general case of nonlinear constituent and constraint functions;
    \item[{\sf (ii)}] assessing the existence of feasible pairs of local parameters and input variables associated with connected nodes; and
    \item[{\sf (iii)}] estimating and updating the search domains of the local feasible set subproblems.
\end{enumerate}
To handle these challenges we propose, respectively:
\begin{itemize}
    \item[\sf (i)] the solution of the subproblems by sampling, with subsequent parameterisation of surrogates $\overline{\mathbf{G}}^{({\sf d})}_{i}$ to describe the coupling constraints;\footnote{The overbar notation in $\overline{\mathbf{G}}^{({\sf d})}_{i}$ is in keeping with Section \ref{sec:prelim} and indicates approximations to the functions evaluated through sampling.}
    \item[\sf (ii)] the solution of auxiliary mathematical programs defined on those surrogates; and
    \item[\sf (iii)] gaining approximations to the search domains through sampling.
\end{itemize} 
\paragraph{Learning to Parameterise Local Feasible Sets and their Relaxations}
With a set of samples and indications of feasibility, $\mathcal{D}^{({\sf d})}_i = \{((\textbf{v}_i, \textbf{u}_i), s)_{k}\}_{k=1}^{K}$, generated in solving a given subproblem through a sampling-based algorithm, one can define the following approximation to a local feasible set,
\begin{equation*}\label{eq:VUapproxF}
    \begin{aligned}
     {\mathbb{VU}}^{({\sf d})}_{i} &\approx \left\{ (\textbf{v}_i, \textbf{u}_i) \in \mathcal{K}^{({\sf d})}_{{vu}_i} ~\middle|~ \textbf{0} \geq \overline{\mathbf{G}}^{({\sf d})}_{i}(\textbf{v}_i, \textbf{u}_i)\right\} , \\
    \end{aligned}    
\end{equation*}
for a suitable $\overline{\mathbf{G}}^{({\sf d})}_{i}:\mathbb{R}^{n_{v_i}}\times\mathbb{R}^{n_{u_i}}\to\mathbb{R}$. As mentioned in Section \ref{sec:BC_SM}, such a parameterisation could be provided by any binary classifier.\footnote{In special cases like in Example \ref{ex:1}, one may be able to identify such a parameterisation without sampling and surrogate learning, given certain conditions on the search domains, forward node evaluations and inequality constraints; whether to use it would also depend on factors such as computational tractability.}

Next, we turn our attention to providing mechanism to couple connected nodes and estimate the search domain for each subproblem.

\paragraph{Forward Relaxation and Propagation}
The forward propagation subproblems are \eqref{eq:DS_unitFP0} are approximated as
\begin{equation}\label{eq:DS_unit5}
    {\mathbb{VU}}^{({\sf f})}_i \approx \left\{ (\textbf{v}_i,\textbf{u}_i) \in \mathcal{K}^{({\sf f})}_{vu_i}~\middle|
    \begin{array}{l}
        \mathbf{0} \geq \mathbf{G}_i(\textbf{v}_i, \textbf{u}_i) \\[2mm]
        \forall j\in\mathcal{N}^{\rm in}_i,\\
        0 \geq \displaystyle\min_{\substack{\textbf{v}_j,\textbf{u}_j \in \mathcal{K}_{vu_{j}}^{({\sf f})}\\ \textbf{v}_k,\textbf{u}_k \in \mathcal{K}^{\sf(f)}_{vu_k}}} \overline{\mathbf{G}}^{({\sf f})}_{j}(\textbf{v}_j, \textbf{u}_j)\\
        \hphantom{\mathbf{0} \geq\ \ }\text{s.t.}\ \textbf{u}_i^j = \overline{\textbf{F}}_{j}^{i}(\textbf{v}_j, \textbf{u}_j)\\
        \quad \quad \quad \quad  \textbf{u}_k^j = \overline{\textbf{F}}_{j}^{k}(\textbf{v}_j, \textbf{u}_j) \\
        \quad \quad \quad \quad {0} \geq \overline{\mathbf{G}}^{({\sf f})}_k(\textbf{v}_k, \textbf{u}_k)
        \\ \quad \quad \quad \quad \forall k \in \mathcal{N}_j^{\rm out}, k\prec i
    \end{array} \right\},
\end{equation}
with $\overline{\textbf{F}}_{j}^{i}$ denoting a surrogate approximation to the forward evaluation $\textbf{F}^i_j$ of node $j$ to $i$.\footnote{Note that one need not always construct an approximation, for example if the constituent function is a simple algebraic expression. However, if the evaluation is the result of a forward solution of a complex model, then building a surrogate rather than embedding the model directly as constraints into the NLP clearly becomes desirable.} The identification of node $j$ surrogates, $\overline{\textbf{G}}^{\sf (f)}_j$ and $\overline{\textbf{F}}_{j}^{i}$ comes at the cost of little computational overhead, since the training data is already available from identification of subproblem $j \prec i$ according to the precedence ordering. The same applies to node $k$ surrogates, $\overline{\textbf{F}}_{j}^{k}$ and $\overline{\textbf{G}}_k^{\sf(f)}$ with $j \prec k \prec i$. The solution to \eqref{eq:DS_unit5} for node $i$ may then be sampled using a nonlinear programming {(NLP)} solver to test for feasibility of candidate pairs $(\textbf{v}_i, \textbf{u}_i)$.

\begin{figure}[htb]
    \centering
    \includegraphics[width=1\linewidth]{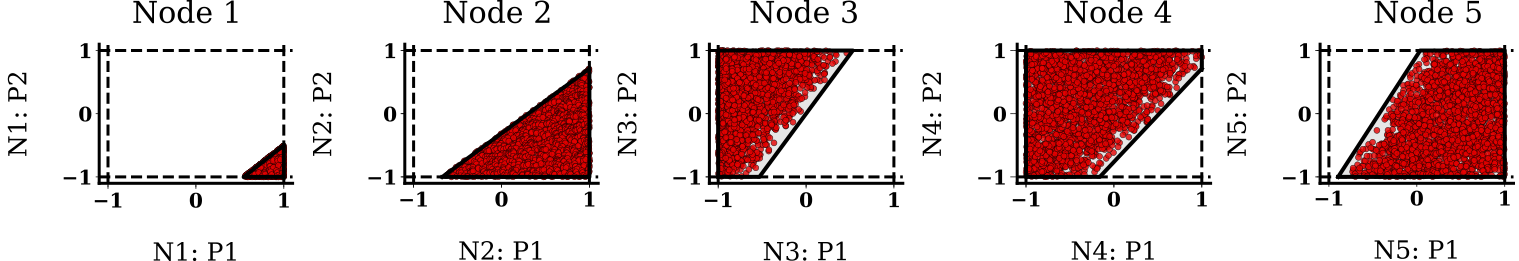}
    \caption{The projections of \eqref{eq:DS_unit5} for Example \ref{ex:1} onto the subspace of node parameters (1--5, left--right) identified through sampling. The projections computed by {\sf MPT 4.0} are indicated by grey shade, with black solid lines indicating the boundary.}
    \label{fig:FPsampled_ex1}
\end{figure}

\begin{continuedexample}{ex:1}
    A solution to each subproblem was sampled using \num{2500} feasible points. The projections onto the subspace of node parameters are shown in Figure \ref{fig:FPsampled_ex1}. Comparison with Figure \ref{fig:approx1_proj} highlights strong similarities in the shape of the recovered sets, with one being continuous and the other discrete. Surrogate constraints were identified via neural network binary classifiers and the affine constituent node functions were approximated using neural network regressors. All surrogates leveraged cross-validation practice for hyperparameter selection. Refer to Section~\ref{sec:accsurr} for further discussion on accuracy of surrogate construction. 
\end{continuedexample}

\paragraph{Backward Relaxation and Propagation}
Likewise, the back-propagation subproblems \eqref{eq:DS_unitBP} are approximated as
\begin{equation}\label{eq:DS_unit4}
    {\mathbb{VU}}^{({\sf b})}_i \approx \left\{ (\textbf{v}_i,\textbf{u}_i) \in \mathcal{K}^{({\sf b})}_{vu_i}~\middle|
    \begin{array}{l}
        \mathbf{0} \geq \mathbf{G}_i(\textbf{v}_i, \textbf{u}_i)\\
        \forall k\in\mathcal{N}^{\rm out}_i,\\
        {0} \geq \displaystyle\min_{\substack{\textbf{v}_k,\textbf{u}_k \in \mathcal{K}^{\sf (b)}_{{vu}_{k}}\\ \textbf{v}_j, \textbf{u}_j \in \mathcal{K}^{\sf (b)}_{vu_j}}}\overline{\mathbf{G}}^{({\sf b})}_k(\textbf{v}_k, \textbf{u}_k)\\
        \hphantom{\mathbf{0} \geq\ \ }\text{s.t.}\ \textbf{u}_k^i = \textbf{F}_i^k(\textbf{v}_i, \textbf{u}_i)\\
        \quad \quad \quad \quad  \textbf{u}_k^j = \overline{\textbf{F}}_j^{k}(\textbf{v}_j, \textbf{u}_j) \\
        \quad \quad \quad \quad {0} \geq \overline{\mathbf{G}}^{({\sf b})}_j(\textbf{v}_j, \textbf{u}_j)
        \\ \quad \quad \quad \quad \forall j \in \mathcal{N}_k^{\rm in}, j\succ i
    \end{array} \right\}.
\end{equation}
The surrogate $\overline{\textbf{G}}^{\sf (b)}_k$ is already available at node $i \prec k$ since the subproblems are solved according to the reverse of the precedence ordering. Note that in the case a box enclosure is selected and constraint {\sf (b)} is dropped from \eqref{eq:DS_unitBP}, the embedded optimisation problem in \eqref{eq:DS_unit4} is a box-constrained {NLP} given that one obtains the inputs to node $k$ from node $i$ from the evaluation of $\textbf{F}_i^k$ reducing the equality constraint into the objective; and all constraints related to nodes $j\in \mathcal{N}_k^{\rm in}$ with $j \succ i$ are removed.

\begin{figure}[htb]
    \centering
    \includegraphics[width=1\linewidth]{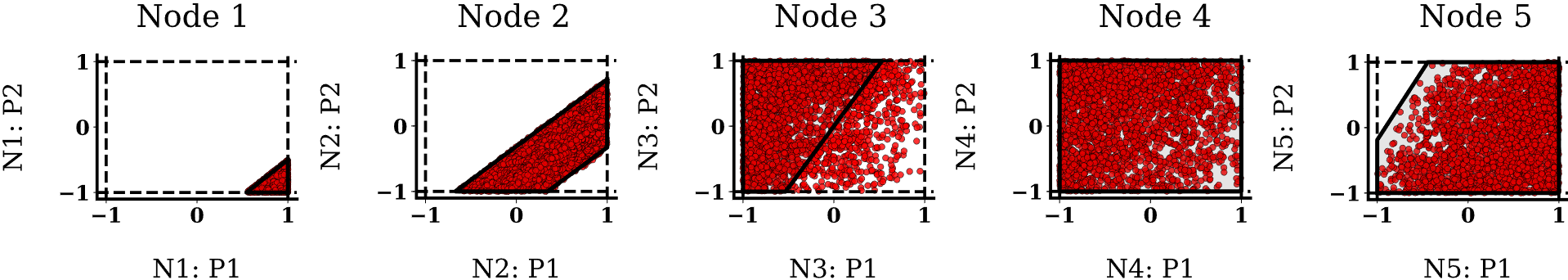}
    \caption{The projections of \eqref{eq:DS_unit4} for Example \ref{ex:1} onto the subspace of node parameters (1--5, left--right) identified through sampling (red). The projections computed by {\sf MPT~4.0} are indicated by grey shade, with black solid lines indicating the boundary.}
    \label{fig:BPsampled_ex1}
\end{figure}

\begin{continuedexample}{ex:1}
    The projections for each subproblem onto the subspace of node parameters for Example \ref{ex:1} are detailed by Figure \ref{fig:BPsampled_ex1}. Each solution used \num{2500} discrete points. Search space initialization utilized a sampling based procedure presented later in this section. In total \num{16384} Sobol samples were used. Surrogates were constructed in the same manner as in the forward propagation, except this time an SVM binary classifier was used for parameterisation of the feasbile set. Comparison with Figure \ref{fig:approx2_proj} again highlights similarities in the shape of recovered sets, one being continuous and the other discrete. However, node 3 is a clear superset of that identified using {\sf MPT 4.0}, which is attributed to the data-driven initialization of $\mathcal{K}_{u_3}$ yielding a larger set than reachable from forward evaluation of nodes $j\prec3$; whereas the search domain $\mathcal{K}_{u_3}$ is accurately calculated using {\sf MPT 4.0}. 
\end{continuedexample}

\paragraph{Composing Propagations}
The recipe for composing propagations follows a similar structure, with the forward-backward propagation subproblems \eqref{eq:DS_unitFBP} approximated as
\begin{equation}\label{eq:DS_unit6}
    {\mathbb{VU}}^{({\sf fb})}_i \approx \left\{ (\textbf{v}_i,\textbf{u}_i) \in \mathcal{K}^{({\sf fb})}_{vu_i}~\middle|
    \begin{array}{l}
        \mathbf{0} \geq \mathbf{G}_i(\textbf{v}_i, \textbf{u}_i) \\[2mm]
        \forall j\in\mathcal{N}^{\rm in}_i,\\
        \mathbf{0} \geq \displaystyle\min_{\substack{\textbf{v}_j,\textbf{u}_j \in \mathcal{K}_{vu_{j}}^{({\sf f})}\\ \textbf{v}_k,\textbf{u}_k \in \mathcal{K}^{\sf(f)}_{vu_k}}} \overline{\mathbf{G}}^{({\sf f})}_{j}(\textbf{v}_j, \textbf{u}_j)\\
        \hphantom{\mathbf{0} \geq\ \ }\text{s.t.}\ \textbf{u}_i^j = \overline{\textbf{F}}_{j}^{i}(\textbf{v}_j, \textbf{u}_j)\\
        \quad \quad \quad \quad  \textbf{u}_k^j = \overline{\textbf{F}}_{j}^{k}(\textbf{v}_j, \textbf{u}_j) \\
        \quad \quad \quad \quad \textbf{0} \geq \overline{\mathbf{G}}^{({\sf f})}_k(\textbf{v}_k, \textbf{u}_k)
        \\ \quad \quad \quad \quad \forall k \in \mathcal{N}_j^{\rm out}, k\prec i \\[2mm]
        \forall k \in \mathcal{N}_i^{\rm out}, \\
        \mathbf{0} \geq \displaystyle\min_{\substack{\textbf{v}_k,\textbf{u}_k \in \mathcal{K}^{\sf (b)}_{{vu}_{k}}\\ \textbf{v}_j, \textbf{u}_j \in \mathcal{K}^{\sf (b)}_{vu_j}}} \overline{\mathbf{G}}^{({\sf fb})}_k(\textbf{v}_k, \textbf{u})\\
        \hphantom{\mathbf{0} \geq\ \ }\text{s.t.}\ \textbf{u}_k^i = \textbf{F}_i^k(\textbf{v}_i, \textbf{u}_i)\\
        \quad \quad \quad \quad  \textbf{u}_k^j = \overline{\textbf{F}}_{j}^{k}(\textbf{v}_j, \textbf{u}_j) \\
        \quad \quad \quad \quad \textbf{0} \geq \overline{\mathbf{G}}^{({\sf fb})}_j(\textbf{v}_j, \textbf{u}_j)
        \\ \quad \quad \quad \quad \forall j \in \mathcal{N}_k^{\rm in}, j\succ i
    \end{array} \right\}.
\end{equation}
A backward-forward composition can be defined analogously.

\paragraph{Defining Search Domains for the Propagations}
Here, discussion is directed towards estimating the sets $\mathcal{K}_{vu_i}^{\sf(d)} = \mathcal{K}_{v_i} \times \mathcal{K}^{\sf (d)}_{u_i}$ with ${\sf d \in \{f,b\}}$ that define the search domain for a node's parameters and input variables. 
Starting with forward propagation, an estimate of $\mathcal{K}^{\sf (f)}_{u_i}$ is required. In the general case, an exact description of $\mathcal{K}_{u_i}^{(\sf f)}$ is unlikely to be obtained, but an enclosure may be computed from samples generated in solving $j \prec i$ as
\begin{equation}
    \mathcal{K}_{u_i}^{\sf (f)}
\approx\mathcal{H}\left(\prod_{j\in\mathcal{N}_i^{\sf \text{in}}}\textbf{F}_j^i[{\mathbb{VU}}_j]\right),
\end{equation}
with $\mathcal{H}$ denoting the interval hull of a set, possibly inflated in each dimension to add conservatism.

Defining the sets $\mathcal{K}_{u_i}^{\sf (b)}$, for the backward propagation is more computationally involved as it entails a separate step prior to solving the initial subproblem. Because there is no closed form for the propagation of the search domains $\mathcal{K}_{v_j}$ through each constituent node function in general, an outer approximation is made to the data generated from sampling the search domains and evaluating the composite function accordingly. In particular, we estimate these sets in one step, by sampling $\mathcal{K}_v$ according to a space filling design and then building interval hulls to the sampled input values of each node $i \in \mathcal{N}$ generated through evaluation of the composite function.

For arbitrary compositions finally, we define the search domain as the interval hull to the sampled set approximation from the previous propagation, for example $\mathcal{K}_{vu_i}^{\sf (fb)} \approx \mathcal{H}\left({\mathbb{VU}}_i^{\sf (f)}\right)$.

The implementation for general setups is reliant on solving {NLP} problems with embedded surrogate constraints and the use of sampling. Since any approximation error present in the surrogate will be propagated between the subproblems, it is imperative that surrogates are validated with high accuracy prior to use in solution to a subsequent subproblem. The identification of global solutions to the defined {NLP} is also desirable for reliable approximation. In case study, we find the identification of local solutions via multi-start schemes to be an acceptable compromise between solution quality and computational expense. 

\paragraph{Reconstructing the Full Feasible Space} 
After propagating the local feasible sets ${\mathbb{VU}}^{({\sf d})}_i$ through each node $i\in\mathcal{N}$ of the {DAG} representative of the composite function, it is immediate to determine an inner-approximation of ${\mathbb{V}}^{\sf (d)}_i$ by projection. The joint feasible region $\mathbb{V}$ may now be defined as in \eqref{eq:DS_joint}, yet with the notable difference that the search domain $\mathcal{K}_v$ can be replaced with the reduced set $\mathbb{V}^{\sf (d)}\subseteq \mathcal{K}_v$, as justified by Property~\ref{prop:monotone_inclusion},
\begin{equation}\label{eq:DS_joint2}
\begin{aligned}
    \mathbb{V} &\approx \left\{ \textbf{v} \in {\mathbb{V}}^{({\sf d})}~\middle|
    \begin{array}{l}
        \forall i\in \mathcal{N},\ \exists (\mathbf{u}_i,\mathbf{y}_i)\in \mathbb{R}^{n_{u_i}}\times\mathbb{R}^{n_{y_i}}:\\
        \mathbf{0} \geq \mathbf{G}_i(\textbf{v}_i, \textbf{u}_i)\\
        \textbf{y}_i^k = \textbf{F}_i^k(\textbf{v}_i, \textbf{u}_i),\ \forall k\in\mathcal{N}^{\rm out}_i\\
        \textbf{u}_k^i = \textbf{y}_i^k,\ \forall k\in\mathcal{N}^{\rm out}_i\\
    \end{array} \right\}. 
\end{aligned}
\end{equation}

Recall our main working hypothesis is that applying a sampling-based algorithm defined on ${\mathbb{V}}^{({\sf d})}$ rather than $\mathcal{K}_v$ will significantly reduce the number of function evaluations. In practice, we often find that simply sampling the discrete Cartesian product set returned from the propagations without adaptation is more efficient than applying adaptive samplers.

\begin{figure}[htbp]
    \centering
    \includegraphics[width=1\linewidth]{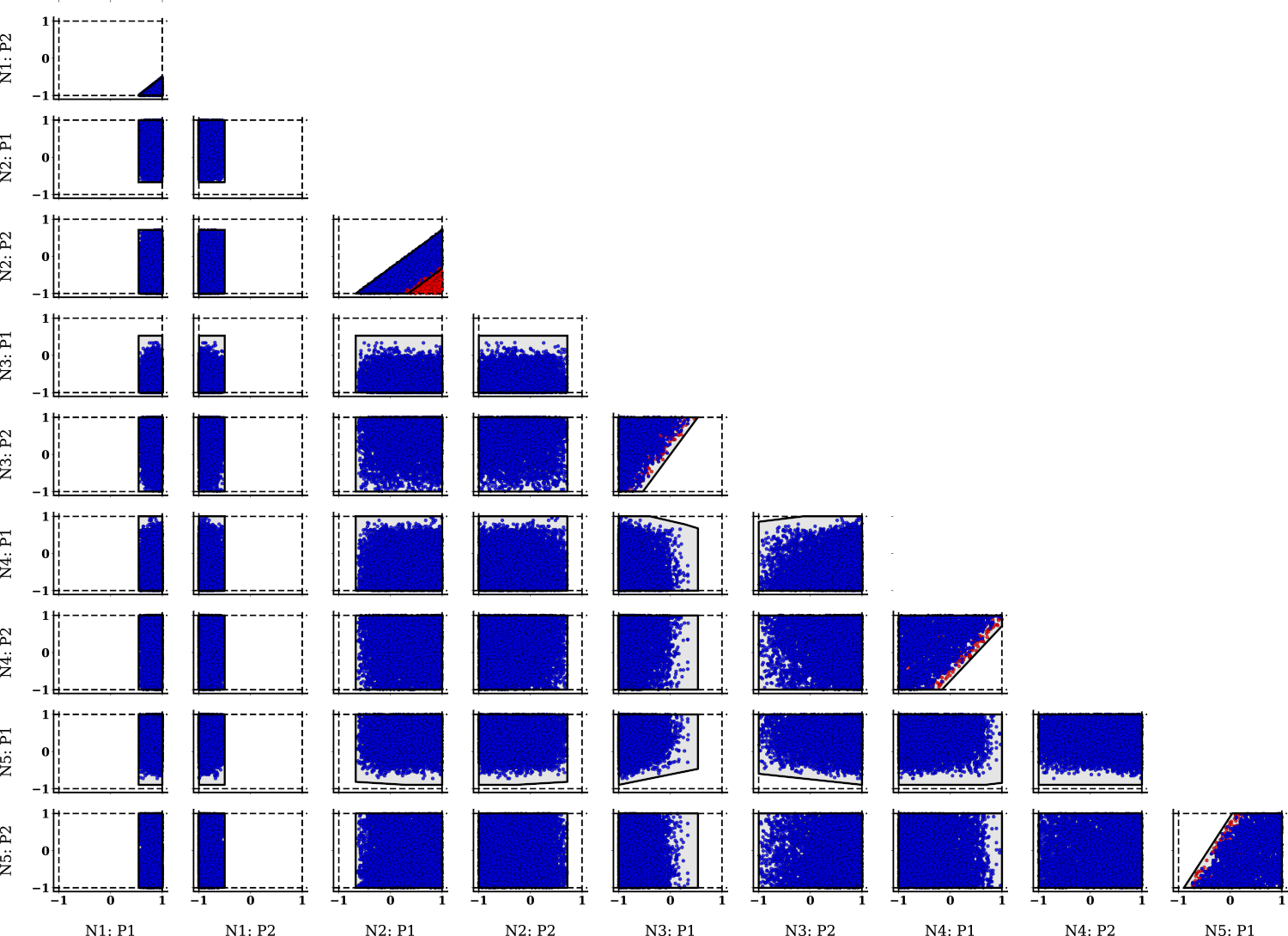}
    \caption{The projections of the reconstructed set from the Cartesian product of forward propagation solutions \eqref{eq:DS_joint2} for Example \ref{ex:1} onto the subspace of node parameters identified through sampling. The light grey shade indicates projections of the solution yielded by {\sf MPT 4.0} with boundary indicated by the black solid lines, the red scatter denotes the solution of the forward propagation, and the blue scatter is the reconstructed set.}
    \label{fig:dsjoint211}
\end{figure}

\begin{continuedexample}{ex:1}
    Figure \ref{fig:dsjoint211} details the reconstructed solution of \num{7000} discrete points (blue scatter) yielded from sampling the Cartesian product of the forward propagation subproblems (each subproblem is identified from \num{2500} points and shown in red scatter). The red shaded regions are the projections of the solution yielded by the {\sf MPT~4.0} solution. It is clear that the sampled solution describes an inner approximation, as is expected from a sampling algorithm. Refer to Section~\ref{sec:sampeff} and Table~\ref{tab:my_table_1} for further discussion on sampling efficiency.
\end{continuedexample} 

\subsection{Extension to Coupling Parameters}\label{sec:extensions}
A key assumption in the formulation of problem \eqref{eq:DS_joint} was that the local constraint and constituent functions  $\textbf{G}_i$ and $\textbf{F}_i$ are dependent only on local parameters $\textbf{v}_i$ that are mutually exclusive with the parameters of other nodes. This section discusses an extension of \eqref{eq:DS_joint} to the case where $\textbf{G}_i$ and $\textbf{F}_i$ are also dependent on coupling variables $\textbf{z} \in \mathcal{K}_z$,
\begin{equation}\label{eq:DS_joint2_aux}
    \mathbb{VZ} \coloneqq \left\{ (\textbf{v}, \textbf{z})\in \mathcal{K}_{vz}~\middle|
    \begin{array}{l}
        \forall i\in \mathcal{N},\ \exists (\mathbf{u}_i,\mathbf{y}_i)\in \mathbb{R}^{n_{u_i}}\times\mathbb{R}^{n_{y_i}}:\\
        \mathbf{G}_i(\textbf{v}_i, \textbf{u}_i, \textbf{z}) \leq \mathbf{0}\\
        \textbf{y}_i = \textbf{F}_i(\textbf{v}_i, \textbf{u}_i,\textbf{z})\\
        \textbf{u}_k^i = \textbf{y}_i^k,\ \forall k\in\mathcal{N}^{\rm out}_i\\
    \end{array} \right\},
\end{equation}
with $\mathcal{K}_{vz} = \mathcal{K}_v \times \mathcal{K}_z$.

\begin{example}
    An intuitive example for the presence of coupling variables is provided by the simultaneous verification and estimation of an approximation $\bar{f}(\textbf{z}, \textbf{v})$ to a given function ${f}(\textbf{z})$ on a compact domain $\mathcal{K}_z \subset \mathbb{R}^{n_z}$. The approximation is described as a composite function and defined by parameters $\textbf{v}\in \mathcal{K}_v$. We wish to identify the subset of parameters $\mathbb{VZ}$ for which the approximator satisfies some maximum point wise error, $\lvert \bar{f}(\textbf{z}, \textbf{v})-f(\textbf{z})\rvert_p \leq \epsilon$ with $\epsilon \in \mathbb{R}_+$. One such case study is investigated in Section \ref{sec:facs}.
\end{example}

The problem \eqref{eq:DS_joint2_aux} can be recast as \eqref{eq:DS_joint} through lifting the coupling parameters \textbf{z} into local copies for each node $i \in \mathcal{N}$ and enforcing their equality between subproblems. Accordingly, an extended local feasible set $\mathbb{VUZ}_i$ corresponding to \eqref{eq:DS_joint2_aux} may be stated as
%
%
\begin{equation}\label{eq:DS_unit2_auxiliary}
    \mathbb{VUZ}_i \coloneqq \left\{ (\textbf{v}_i,\textbf{u}_i, \textbf{z}_i) \in \mathcal{K}_{vuz_i} \middle|
    \begin{array}{l}
        \mathbf{0} \geq \mathbf{G}_i(\textbf{v}_i, \textbf{u}_i, \textbf{z}_i) \\[2mm]
        \forall k\in\mathcal{N}^{\rm out}_i,\ \exists (\textbf{v}_k,\textbf{u}_k, \textbf{z}_k) \in \mathbb{VUZ}_k:\\
        \left[\textbf{u}_k^i, \textbf{z}_k\right] = \left[{\textbf{F}}_i^k(\textbf{v}_i, \textbf{u}_i, \textbf{z}_i), \textbf{z}_i\right] \\[2mm]
        \forall j\in\mathcal{N}^{\rm in}_i,\ \exists (\textbf{v}_j,\textbf{u}_j, \textbf{z}_j) \in \mathbb{VUZ}_j:\\
        \big[\textbf{u}_i^j, \textbf{z}_i\big] = \left[{\textbf{F}}_j^i(\textbf{v}_j, \textbf{u}_j, \textbf{z}_j), \textbf{z}_j\right]\\
    \end{array} \right\}\,,
\end{equation}
with $\mathcal{K}_{vuz_i} = \mathcal{K}_{v_i} \times \mathcal{K}_{u_i} \times \mathcal{K}_z$. The extended constituent functions $[{\textbf{F}}_i^k(\textbf{v}_i, \textbf{u}_i, \textbf{z}_i), \textbf{z}_i]$ and variables $[\textbf{u}_k^i, \textbf{z}_k]$ enable propagation of the lifted parameters $\textbf{z}_i$ between the nodes.

The following properties follow readily from recasting \eqref{eq:DS_joint2_aux} as \eqref{eq:DS_joint} using parameter lifts:
\begin{property}\label{prop:ps_decom}
    The problem structure provided by \eqref{eq:DS_unit2_auxiliary} admits relaxations and propagations similar to those of \eqref{eq:DS_unit2} with appropriate modification to the definitions of \eqref{eq:DS_unitFP0}, \eqref{eq:DS_unitBP} or \eqref{eq:DS_unitFBP}. Subproblems are denoted as ${\mathbb{VUZ}}_i^{\sf (d)}$ with ${\sf d \in \{ f,b, fb, bf, \ldots\}}$.
\end{property}
\begin{property}\label{prop:aux}
    The reformulation provided by \eqref{eq:DS_unit2_auxiliary} implies that the projection of the extended local feasible set onto the subspace of lifted parameters is equal for all nodes $i \in \mathcal{N}$,
    \begin{equation*}
        \mathbb{Z}_i = \left\{ \textbf{z}_i \in \mathcal{X}~\middle|
    \begin{array}{l}
        \exists (\textbf{v}_i,\textbf{u}_i) \in \mathcal{K}_{{vu}_i}:
        (\textbf{v}_i,\textbf{u}_i, \textbf{z}_i) \in \mathbb{VUZ}_i
    \end{array} \right\} \eqqcolon \mathbb{Z},
    \end{equation*}
    with $\mathbb{Z}$ referred to as the reduced domain.
\end{property}
\begin{property}\label{prop:reduce_f_b}
   A reduced search domain $\mathbb{VZ}^{\sf (d)}$ for a simultaneous approach to \eqref{eq:DS_joint2_aux} is given by the Cartesian product of the projections of the propagations onto the subspace of both local and coupling parameters:
    \begin{equation*}
        \forall {\sf d \in \{f, b, fb, bf, \ldots\}},\quad {\mathbb{VZ}}^{\sf (d)} \coloneqq \prod_{i\in \mathcal{N}}{\mathbb{V}}^{\sf (d)}_i \times \bigcap_{i\in \mathcal{N}} \mathbb{Z}_i^{\sf (d)} \supseteq \mathbb{VZ}.
    \end{equation*}
\end{property}

Overall, Property~\ref{prop:monotone_inclusion} holds as before through lifting any coupling parameters as in \eqref{eq:DS_joint2_aux}, preserving the monotone inclusion sequence of sets obtained through alternating composition of forward and backward propagations and their convergence.

\section{Software and Implementation}\label{sec:software_imp}

The following case study results were generated by the open-source python software $\boldsymbol{\mu}\textbf{F}$, which is general to arbitrary composite functions and constraint definitions considered in the paper. The software automates
\begin{itemize}
    \item the decomposition, relaxations, propagations of local feasible sets and their composition, as well as the reconstruction of the full feasible set;
    \item the evaluation of constraints and forward node evaluations;
    \item the training and hyperparameter tuning of surrogates through \textit{k}-fold cross validation;
    \item the formulation and solution of the embedded NLP problems.
\end{itemize}
All that is required from the user is the configuration of hyperparameters (as handled by {\sf Hydra} \citep{Yadan2019Hydra}), selection of the sampler, and definition of the constraint and constituent functions. All relevant objects are stored within a {DAG} structure provided by {\sf NetworkX} \citep{hagberg2008exploring}. The package currently interfaces the nested sampling schemes detailed in the python package {\sf DEUS} by default.\footnote{\ifanon Repository URLs withheld for review.\else$\boldsymbol{\mu}\textbf{F}$: \url{https://github.com/mawbray/mu.F}\newline {\sf DEUS}: \url{https://github.com/omega-icl/deus}\fi} A general interface is also provided for a user to select an adaptive sampler of choice. The package leverages {\sf JAX} \citep{jax2018github} for just-in-time compilation, automatic differentiation, and use of accelerators. Support for training of surrogates, for example SVM, neural networks and Gaussian process, utilizes {\sf Flax}, {\sf GPJax} and {\sf Scikit-Learn}. {\sf JAXOpt} \citep{jaxopt_implicit_diff} is used to solve box-constrained {NLP}s via the Limited-memory Broyden-Fletcher-Goldfarg-Shanno-B ({\sf L-BFGS-B}) algorithm \citep{zhu1997algorithm}. At the current time, no {\sf JAX} packages interface with state-of-the-art general {NLP} solvers. Therefore, {\sf CasADi} \citep{Andersson2019} is leveraged for solution of {NLP}s with general constraints. Specifically, we utilize {\sf CasADi} callbacks to embed {\sf JAX}-based surrogate models into an {NLP}, and then pass to a solver such as {\sf IPOPT} \citep{wachter2006implementation}. This also allows {\sf JAX} to supply the relevant Jacobian and Hessian information through automatic differentiation, without hard coding the surrogate structure via {\sf CasADi} symbolics. Given the use of local solvers, a multi-start scheme is used to handle nonconvexity in the auxiliary {NLP}s. To take advantage of the parallelism of sampling and to ensure the safety of the solver operations, the {NLP} solution utilizes process-based parallelism through {\sf Ray} \citep{moritz2018ray}. 

\section{Case Studies}\label{sec:CS}
In addition to Example \ref{ex:1} used to illustrate the steps in the methodology throughout Section~\ref{sec:method}, we present three case studies. The first two have relevance to industrial process operation. They explore identification of feasible operational parameters, respectively, in chemical production and tableting operation in pharmaceutical manufacturing. The final case study entails the simultaneous estimation and verification of a function approximation.\footnote{All study details and results can be found at $\boldsymbol{\mu}\textbf{F}$.} The results of the case studies are first presented qualitatively in Sections~\ref{sec:reacnet}--\ref{sec:facs}, before analysing them quantitatively in terms of surrogate modelling accuracy and sample efficiency in Section~\ref{sec:CSquant}.

Key properties of the case studies are summarized by Table \ref{table:cs_summary}, according to the order of presentation. Each study is associated with a solver; classes of constituent functions including algebraic {\sf (Alg)}, transcendental {\sf (Tran)}, and ordinary differential equations {\sf (ODE)}; affine or nonlinear nature of the constraints in terms of local/coupling parameters and input variables; number $N$ of nodes within the evaluation {DAG}; dimensionality $(n_{v_i} + n_{u_i} + n_z)$ of the extended feasible space subproblems; and dimensionality $n_{v}$ of the full problem. 

\begin{table}[htb]
\begin{center}
\caption{Key properties of the case studies explored. }\label{table:cs_summary}
\footnotesize
\begin{tabular}{ccccccc}
\toprule

\# & Solver & Functions & Constraints & $\lvert \mathcal{N} \rvert$ & ($n_{v_i} + n_{u_i} + n_z$) & $n_v$ \\
\midrule 
1 & {\sf MPT 4.0 / DEUS} & {\sf Alg} & Affine & $5$ & $(2,2,4,3,3)$ & $10$ \\ 
2 & {\sf DEUS} & {\sf ODE} & Nonlinear & $2$ & $(2,4)$ & $4$ \\
3 & {\sf DEUS} & {\sf Alg + Tran} & Nonlinear & $3$ & $(2,4,3)$ & $7$ \\
4 & {\sf DEUS} & {\sf Alg + Tran} & Nonlinear & $6$ & $(4, 5, 5, 5, 6, 8)$ & $13$ \\
\bottomrule
\end{tabular}
\end{center}
\end{table}
\subsection{Serial Batch Reactor Network}\label{sec:reacnet}
\subsubsection{Problem Statement}
Consider a chemical process that consists of two batch reactors connected in series.
\begin{center}
    \begin{tikzpicture}
    \node[draw, circle] (1) at (0,0) {1};
    \node[draw, circle] (2) at (2,0) {2};
    \draw[->] (1) -- (2);
    \end{tikzpicture}
\end{center}
The reaction mechanism characterising the two reactors is the same,
\begin{equation*}
    2{\sf A} \xrightarrow{k_1} {\sf B} \xrightarrow{k_2} {\sf C}\,,
\end{equation*}
where component {\sf B} is the desired product and {\sf C} is an (inert) byproduct.
The reactor dynamics are ODEs that describe the evolution of component molar concentrations in continuous time. The system adheres to Arrhenius kinetics,
\begin{equation*}
    k_{r,j}(T_j) = k^\circ_{r,j} \exp\Bigg(-\frac{E_{r,j}}{RT_j}\Bigg)\,,
\end{equation*}
where $T_j$ is the temperature (K) of reactor $j\in \mathcal{N}$; $k_{r,j}$, the temperature-dependent rate constant of reaction step $r\in\{1,2\}$; $k^\circ_{r,j}$, the pre-exponential factor; $E_{r,j}$, the activation energy; and $R$ denotes the ideal gas constant. The kinetic parameters detailed in Table~\ref{table:kinetics} differ for the two reactors but are assumed to be known with certainty. The volume of both reactors is set to $V=1~\text{m}^3$. 
\begin{table}[tb]
\begin{center}
\caption{Arrhenius kinetic parameter values used in the reaction network.}\label{table:kinetics}
\small
\begin{tabular}{ccccc}
\toprule
 \multirow{2}{*}{Parameter} & \multicolumn{2}{c}{Reactor 1} & \multicolumn{2}{c}{Reactor 2} \\
\cmidrule{2-5}
  & Reaction 1 & Reaction 2 & Reaction 1 & Reaction 2 \\
\midrule 
$k^\circ_r\text{       (m}^3\: \text{kmol}^{-1} \text{min}^{-1})$ & $1.66 \times 10^{-4}$  & $0.50$ & $9.66 \times 10^{-3}$  & $5.03$ \\ \
$E_r \text{       (kJ\: mol}^{-1})$ & $1.50$ & $5.00$ & $2.50$ & $5.00$\\
\bottomrule
\end{tabular}
\end{center}
\end{table}
Each {ODE} system is given by
\begin{align*}
        \dot{c}_{{\sf A},j}(t) &= -2k_{1,j}(T_j)\,c_{{\sf A},j}(t)^2\\ 
        \dot{c}_{{\sf B},j}(t) &= k_{1,j}(T_j)\,c_{{\sf A},j}(t)^2 - k_{2,j}(T_j)\,c_{{\sf B},j}(t)\\ 
        \dot{c}_{{\sf C},j}(t) &= k_{2,j}(T_j)\,c_{{\sf B},j}(t)\,,
\end{align*}
where $c_{i,j}(t)$ ($\text{kmol m}^{-3}$) denotes the molar concentration of component $i$ in reactor $j$. Each reactor is operated over a time horizon $[0, \tau_j]$, where $\tau_j$ is the batch time. The initial concentration of the first reactor is $\textbf{c}_1(0)=[2\  0\ 0]^\intercal$. The initial concentration of the second reactor is $\textbf{c}_2(0)=[c_{{\sf A},1}(\tau_1)\ c_{{\sf B},1}(\tau_1)\ 0]^\intercal$, with $\textbf{y}_1= \textbf{u}_2= [c_{{\sf A},1}(\tau_1)\ c_{{\sf B},1}(\tau_1)]^\intercal$, which is identical to the composition present in the first reactor at batch endpoint $\tau_1$, subject to a separation step to completely remove component {\sf C}. The connecting stream is such that $\textbf{y}_1^2 = \textbf{u}_2^1 = [c_{{\sf A},1}(\tau_1)\ c_{{\sf B},1}(\tau_1)]^\intercal$. 

The parameters of interest in the characterisation are the temperature and batch time of both reactors, such that $\textbf{v}_i = [\tau_i\ T_i]^\intercal$. These parameters determine the operational strategy of the system and their feasible settings determine set-points for a control scheme. Inequality constraints are imposed to ensure the efficacy of the separation and the molar purity of component {\sf B} at the end of operating the second batch reactor,
\begin{equation*}
\begin{aligned}
    \frac{c_{{\sf C},1}(\tau_1)}{c_{{\sf A},1}(\tau_1) + c_{{\sf B},1}(\tau_1) + c_{{\sf C},1}(\tau_1)} \leq\ & \eta^{\rm up}_{{\sf C},1}\\
    \frac{c_{{\sf B},2}(\tau_2)}{c_{{\sf A},2}(\tau_2) + c_{{\sf B},2}(\tau_2) + c_{{\sf C},2}(\tau_2)} \geq\ & \eta^{\rm up}_{{\sf B},2}\,,
\end{aligned}
\end{equation*}
with composition targets of $\eta^{\rm up}_{{\sf C},1}\ = 0.240$ and $\eta^{\rm up}_{{\sf B},2}=0.825$. The search domains are set to $(\tau_i, T_i) \in \mathcal{K}_{v_i} = [300,900] \times [300, 700]$ for $i =1,2$. 

\subsubsection{Results of the Decomposition Strategy}
\begin{figure}[tb]
    \centering
    \includegraphics[width=0.485\linewidth]{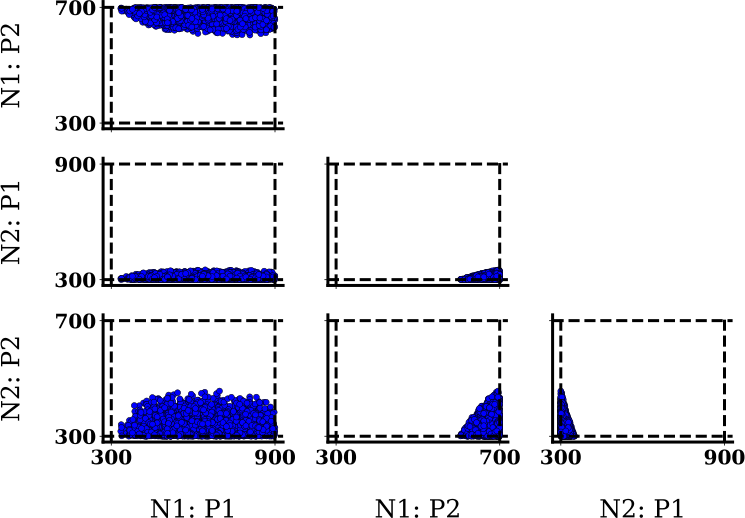}
    \caption{Corner plot of the joint feasible region of operational parameters within the reaction network yielded from the simultaneous approach. Each subplot details the projection of the sampled joint feasible set ${\mathbb{V}}$ onto pairs of process parameters, with dashed lines corresponding to the boundary of $\mathcal{K}_v$.}
    \label{fig:seqDS}
\end{figure}

The following analysis investigates the behavior of the method when a single backward propagation is applied, and then the joint feasible region of operational parameters ($\tau_1, T_1, \tau_2, T_2$) is reconstructed according to the methodology detailed in Section \ref{sec:decomp}. For comparison the full feasible region yielded from the simultaneous approach is visualized by Figure \ref{fig:seqDS}.

\begin{figure}[tb]
    \centering
    \begin{subfigure}[b]{0.485\textwidth}
        \centering
        \includegraphics[width=\textwidth]{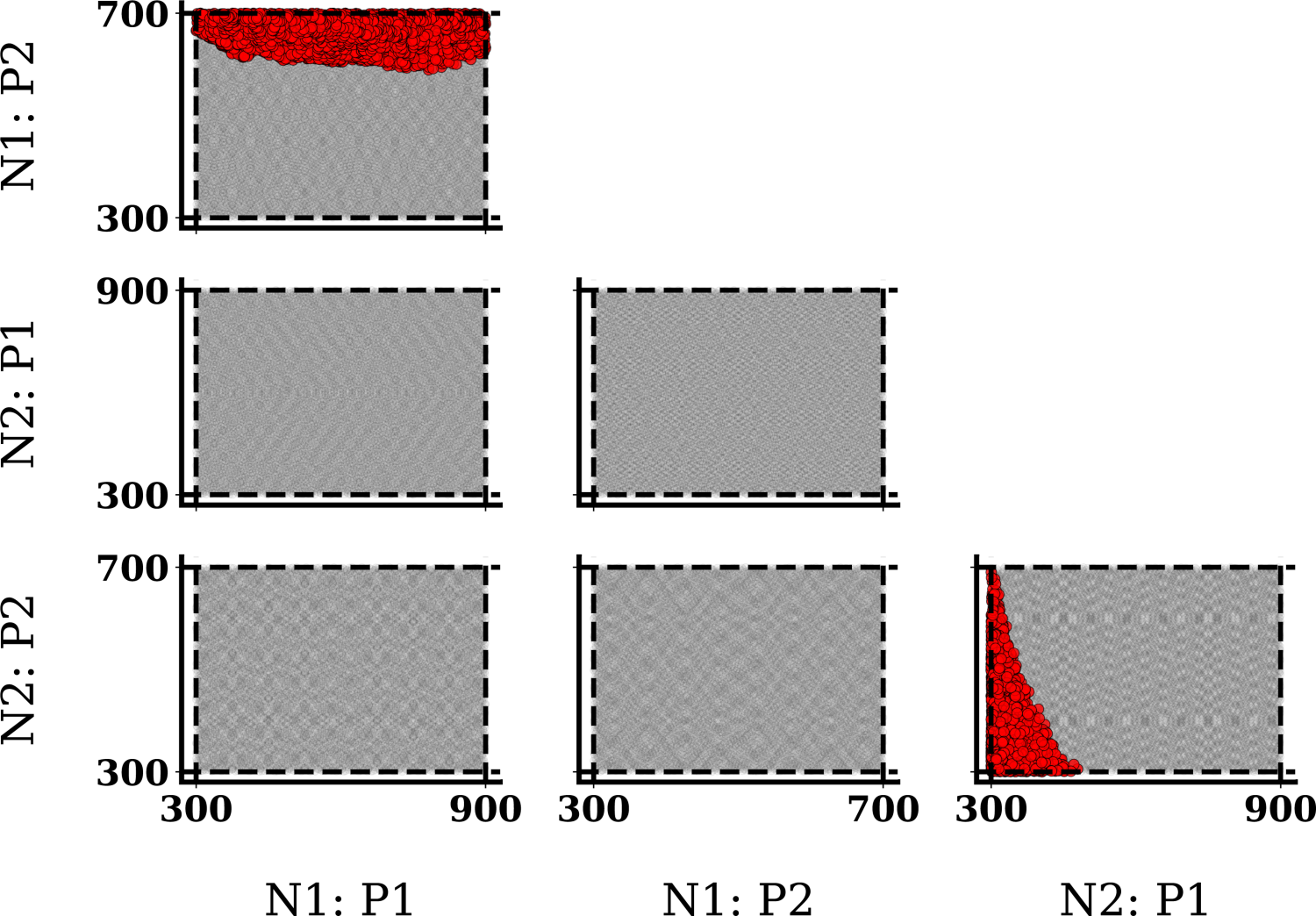}
        \caption{Results of backward propagation}
        \label{fig:sub1}
    \end{subfigure}
    \hfill
    \begin{subfigure}[b]{0.485\textwidth}
        \centering
        \includegraphics[width=\textwidth]{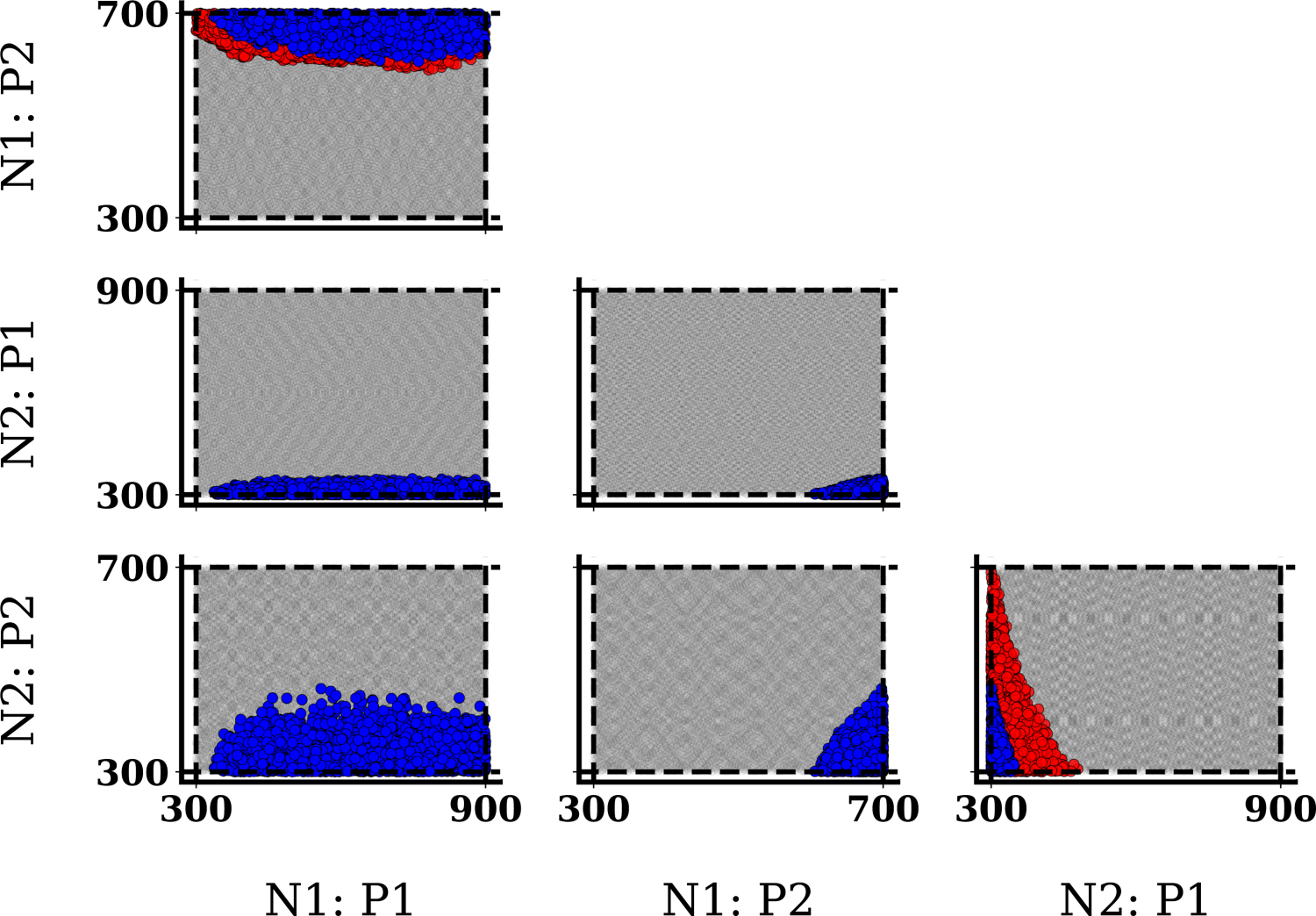}
        \caption{Results of the reconstruction.}
        \label{fig:sub2}
    \end{subfigure}
    \caption{Corner plots of the joint feasible region of operational parameters for the reactor network at various stages of the decomposition methodology. (a) The backward propagation results in red scatter, with samples utilized in initialization of the search space in grey scatter. (b) The reconstruction of the joint feasible set in blue scatter, with the backward propagation in red scatter. The dashed lines define the boundary of the projections of $\mathcal{K}_v$ onto the subspace of pairwise parameters.}
    \label{fig:main_BRN}
\end{figure}

The problem decomposes unit-wise via the proposed approach into two subproblems, with $1\prec 2$. Initialization of the search domains $\mathcal{K}^{\sf (b)}_{u_1}$ and $\mathcal{K}^{\sf (b)}_{u_2}$ leveraged forward simulation of the composite function representative of the process network given $8192$ Sobol samples drawn from $\mathcal{K}_v$. An outer-approximation to the node input variable data generated was then identified as a box enclosure, where the range of the data in each dimension was inflated by 5\%. The samples are plotted in Figure~\ref{fig:sub1} (grey scatter).

Starting from reactor 2, the subproblem ${\mathbb{VU}}^{\sf (b)}_2$ is constituted by four variables $(\textbf{v}_2,\textbf{u}_2^1)\in\mathbb{R}^2\times\mathbb{R}^2$. An inner-approximation with 3500 samples ${\mathbb{VU}}^{\sf (b)}_2$ was determined by nested sampling and the projection on to the subspace of local parameters ${\mathbb{V}}^{\sf (b)}_2$ is demonstrated in the bottom right subplot of Figure~\ref{fig:sub1} (red scatter).

Moving backward to reactor 1, the SVM classifier $\overline{\textbf{G}}^{\sf(b)}_2$ was trained to couple ${\mathbb{VU}}^{\sf (b)}_2$ to the subproblem ${\mathbb{VU}}^{\sf (b)}_1$ defined on reactor 1 upstream. A 2-fold cross-validation strategy was used for hyperparameter selection. The subproblem ${\mathbb{VU}}^{\sf (b)}_1$ comprises two variables $\textbf{v}_1\in\mathbb{R}^2$ and no input variables. An inner-approximation of $\mathbb{VU}_1^{\sf(b)}=\mathbb{V}_1^{\sf(b)}$ with 3500 samples identified through nested sampling is visualized by the top plot of Figure~\ref{fig:sub1}.

The final step entailed reconstructing the joint space ${\mathbb{V}}$ according to a uniform sampling policy defined on the discrete Cartesian product ${\mathbb{V}}^{\sf (b)} \coloneqq {\mathbb{V}}^{\sf (b)}_1\times {\mathbb{V}}^{\sf (b)}_2$ as provided by the backward propagation with 3500 parameter values. This final result of the decomposition approach is detailed on Figure~\ref{fig:sub2} (blue scatter).

The volume of the joint feasible region is suitably recovered by the reconstruction, indicating that the decomposition is effective. This is highlighted through comparisons of the projected sets in Figure~\ref{fig:sub2} (blue scatter) and a realization of the set ${\mathbb{V}}$ obtained via the simultaneous approach in Figure~\ref{fig:seqDS}. Figure~\ref{fig:sub2} additionally demonstrates that the backward relaxation and propagation identifies a superset to the local feasible sets, ${\mathbb{VU}}^{\sf (b)}_i\supseteq \mathbb{VU}_i$. It is worth noting that there is far less overestimation associated to the approximation of reactor 1. This overestimation is attributable to approximation error introduced through sampling.
\FloatBarrier
\subsection{Tableting Operation}

\subsubsection{Problem Statement}
\begin{figure}
    \centering
    \includegraphics[width=0.34\linewidth]{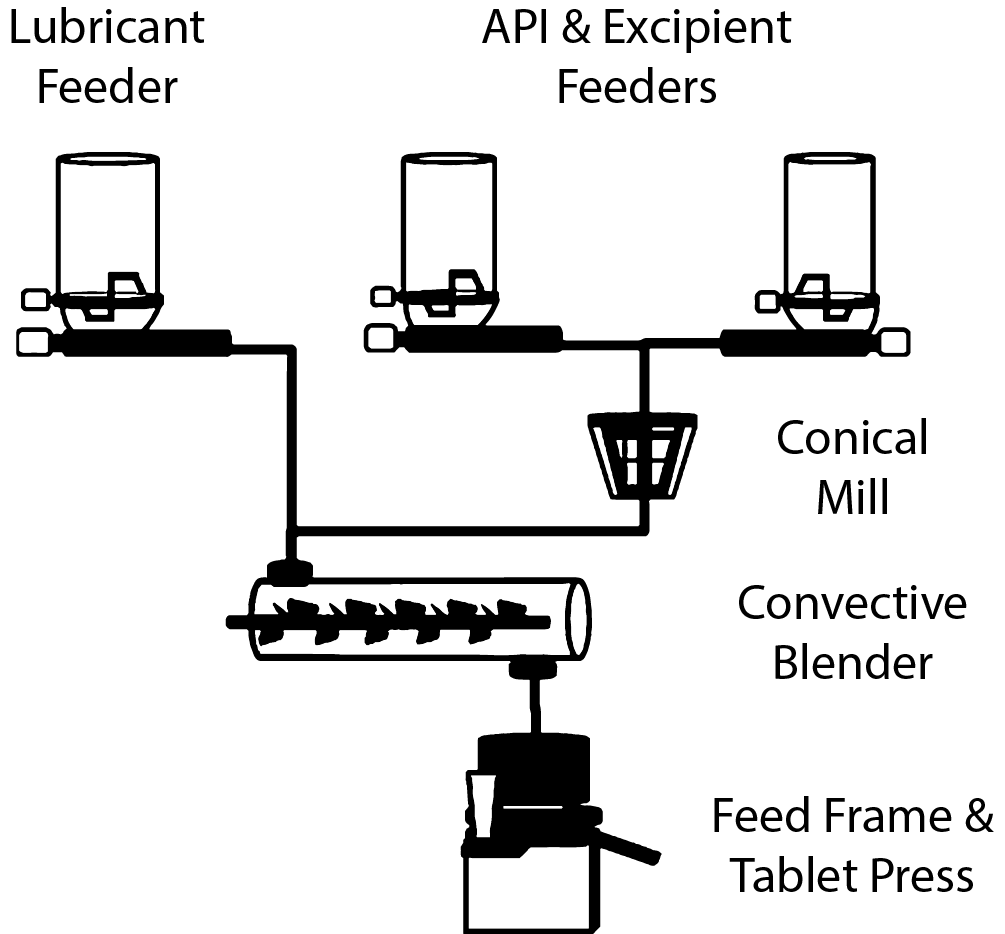}
    \caption{Drug tablet production network \citep{geremia2024design}}
    \label{fig:direct_compaction}
\end{figure}

Consider a multi-unit production network composed of three main unit operations: {\sf1)} co-mill, {\sf2)} convective blender, and {\sf3)} feed frame and tablet press \citep{geremia2024design}. The joint operation of these units must produce drug tablets with suitable quality attributes. The co-mill operation aims to ensure de-lumping of the excipient and active pharmaceutical ingredient {(API)} composite. The blender is operated to ensure homogeneity and a lubricant is added prior within the network structure to ensure flowability and {API} mass fraction within the tablets. The tablet press is then operated to ensure tablet hardness and size. The evaluation graph is given below
\begin{center}
    \begin{tikzpicture}
    \node[draw, circle] (1) at (0,0) {1};
    \node[draw, circle] (2) at (2,0) {2};
    \node[draw, circle] (3) at (4,0) {3};
    \draw[->] (1) -- (2);
    \draw[->] (2) -- (3);
    \end{tikzpicture}
\end{center}
and provides a representation of the processing network defined in Figure~\ref{fig:direct_compaction}. 

As before, we are interested in determining a set of feasible operational parameter settings. A concise description of the unit models is provided next. 
\paragraph{Node 1: Co-Mill}
The parameters of interest are $\textbf{v}_1 = [{\sf F}_{\text{in}}^{\text{api}}\ {\sf F}_{\text{in}}^{\text{exc}}\ {\sf R_{cm}}]^\intercal$, denoting the inflow rates of {API} and excipient and the co-mill blade speed, respectively. There are no input variables to consider. The mass flowrate into the co-mill, ${\sf F}_{\text{in}}^{\text{cm}}$ is the sum of ${\sf F}_{\text{in}}^{\text{api}}$ and ${\sf F}_{\text{in}}^{\text{exc}}$. At steady state, $\sf F^{cm}_{in}$ is equal to the mass flowrate out of the co-mill, ${\sf F}_{\text{out}}^{\text{cm}}$. The mass hold-up within the co-mill at steady state, ${\sf M}_{\text{ss}}^{\text{cm}}$ is nonlinearly related to the parameters $\textbf{v}_1$ via algebraic and transcendental expressions. The bulk density within the co-mill also varies as a nonlinear algebraic function of $\textbf{v}_1$, identified through response surface methodology. One process constraints places an upper bound, ${\sf V}^{\rm up}_{\text{cm}}$ on the volume of the mass hold-up within the co-mill, 
\begin{equation*}\label{eq:cons_1}
\begin{aligned}
    \frac{{\sf M}_{\text{ss}}^{\text{cm}}}{\rho_{\text{bulk}}^{\text{cm}}} \leq  {\sf V}^{\rm up}_{\text{cm}}\,. 
\end{aligned}
\end{equation*}
The co-mill provides the stream $\textbf{y}^2_1 = [{\sf F}_{\text{out}}^{\text{api}}\ {\sf F}_{\text{out}}^{\text{exc}}]^\intercal$ to the convective blender downstream, which consists of the mass flowrate of {\sf API} and excipient out of the unit.
\paragraph{Node 2: Convective Blender}
The parameters of interest are $\textbf{v}_2 = [{\sf F}_{\text{in}}^{\text{lub}}\ {\sf R_{bdr}}]^\intercal$, denoting the mass inflow rate of lubricant and the blender speed, respectively. The node input variables $\textbf{u}^1_2 = [{\sf F}_{\text{out}}^{\text{api}}\ {\sf F}_{\text{out}}^{\text{exc}}]^\intercal$ are the respective mass outflow rates of {API} and excipient from the co-mill. The mass flowrate into the blender, ${\sf F}_{\text{in}}^{\text{bdr}}$ is the sum of ${\sf F}_{\text{out}}^{\text{cm}}$ and ${\sf F}_{\text{in}}^{\text{lub}}$ and is equal to the mass flowrate out of the blender, ${\sf F}_{\text{out}}^{\text{bdr}}$ under steady-state operation. The mass held up within the blender, ${\sf M}^{\text{bdr}}_{\text{ss}}$ is related to the node parameters $\textbf{v}_2$ and input variables $\textbf{u}^1_2$ via a known nonlinear algebraic expression. Similarly, the bulk density of the powder, ${\rho}^{\text{bdr}}_{\text{bulk}}$ and the particle density, $\rho_{\text{p}}^{\text{bdr}}$ within the blender are related to the node parameters and variables through nonlinear algebraic expressions. This provides means to estimate the porosity, $\epsilon_0$ of the blender outlet prior to tableting, i.e. the uncompressed powder porosity,
\begin{equation}\label{eq:pre_comp_porosity}
    {\epsilon_0} = 1 - \frac{{\rho}^{\text{bdr}}_{\text{bulk}}}{\rho_{\text{p}}^{\text{bdr}}}\,.
\end{equation}
One process constraint places an upper bound, ${\sf V}^{\rm up}_{\text{bdr}}$ on the volume hold-up within the blender,
\begin{equation*}\label{eq:cons_2}
\begin{aligned}
   \frac{{\sf M}_{\text{ss}}^{\text{bdr}}}{\rho_{\text{bulk}}^{\text{bdr}}} \leq {\sf V}^{\rm up}_{\text{bdr}}\,.
\end{aligned}
\end{equation*}
Bound constraints are also imposed on the mass fraction of the {API}, ${\sf m}_{\rm api}= \frac{{\sf F}^{\rm api}_{\rm out}}{{\sf F}^{\rm bdr}_{\rm out}}$ within the composition of the tablet,
\begin{equation*}
    {\sf m}^{\rm lo}_{\text{api}} \leq {\sf m}_{\text{api}} \leq {\sf m}^{\rm up}_{\text{api}}\,.
\end{equation*}
The blender provides the stream $y_2^3 = [\epsilon_0]$ to the feed frame and tablet press. 

\paragraph{Node 3: Feed Frame and Tablet Press}
The operation of previous units within the process network dictates the pre-compression porosity $\epsilon_0$ of the tablets, as defined in \eqref{eq:pre_comp_porosity}. The critical process parameters of the feed frame and tablet press $\textbf{v}_{3} = [{\sf P}_{\text{pc}}\ {\sf P}_{\text{main}}]^\intercal$ are deemed to be the pre-compression and the main compression pressure, respectively. These pressures define the operation of two distinct phases: {\sf (i)} a pre-compression step, and {\sf (ii)} a main compression step. The operation of these two phases together with the initial powder volume, $\sf V_0$ determines the pre-compression volume of the powder within the die, ${\sf V}_{\rm pre}$ and the final tablet volume, ${\sf V}_{\rm main}$, which are nonlinearly related to $\epsilon_0$ as well as ${\sf P}_{\text{pc}}$ and ${\sf P}_{\text{main}}.$ The ${\sf P}_{\text{pc}}$ and ${\sf P}_{\text{main}}$ specifications also effect the tablet hardness, ${\sf H}$ through algebraic and transcendental expressions. Bound constraints are imposed on the hardness and tickness of the tablet produced,
\begin{align*}
    {\sf H}^{\rm lo} \leq\ & {\sf H} \leq {\sf H}^{\rm up}\\
    {\sf T}^{\rm lo} \leq\ & \frac{\sf V_{main}}{{\sf A}_d} \leq {\sf T}^{\rm up}\,,
\end{align*}
where ${\sf A}_d$ is the area of the die cross-section.
Full definition of the expressions as well as upper and lower bounds constituting constraints are provided by the $\boldsymbol{\mu}\textbf{F}$ repository. 

\begin{figure}[tb]
    \centering
    \includegraphics[width=0.78\linewidth]{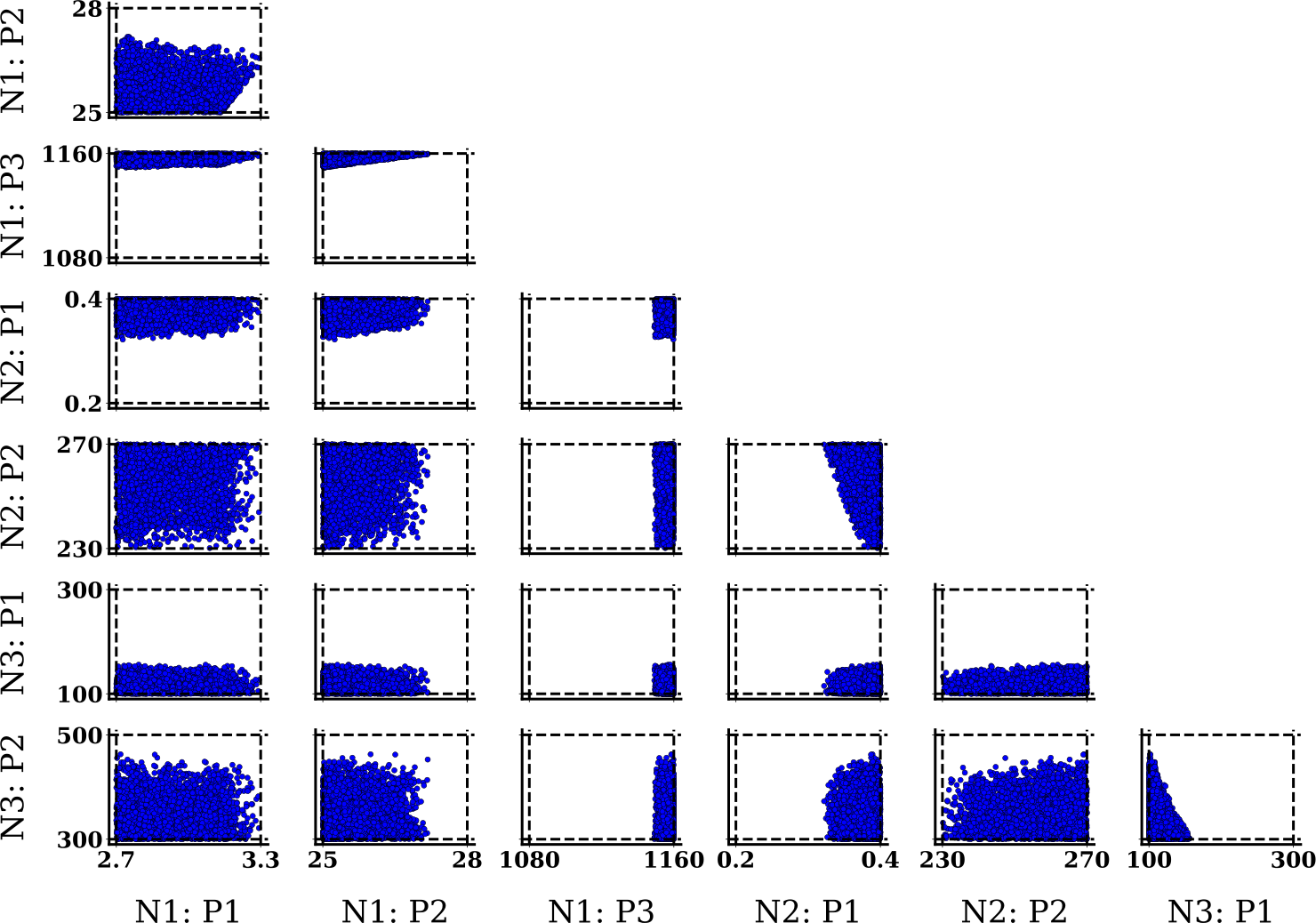}
    \caption{Corner plot of the joint feasible region of operation within the tablet production network yielded from the simultaneous approach.. Each subplot details the projection of the sampled joint feasible set, ${\mathbb{V}}$ onto pairs of process parameters, with the dashed black lines corresponding to the boundary of $\mathcal{K}_v$.}
    \label{fig:directable}
\end{figure}
\begin{figure}[tbp]
    \centering
    \begin{subfigure}[b]{1\textwidth}
        \centering
        \includegraphics[width=0.7\textwidth]{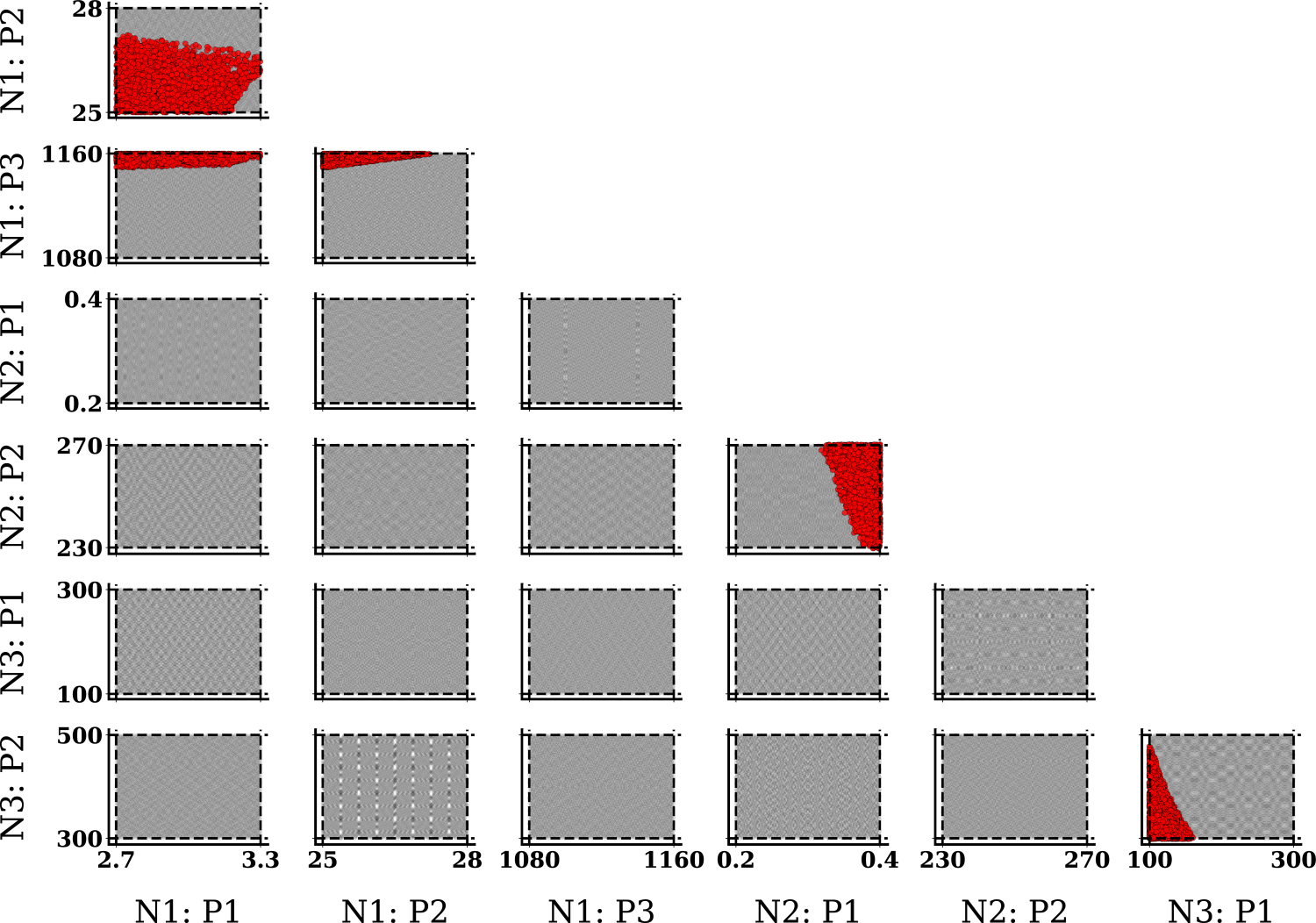}
        \caption{Results of the backward-forward propagation.}
        \label{fig:sub1_tablet}
    \end{subfigure}

    \begin{subfigure}[b]{1\textwidth}
        \centering
        \includegraphics[width=0.7\textwidth]{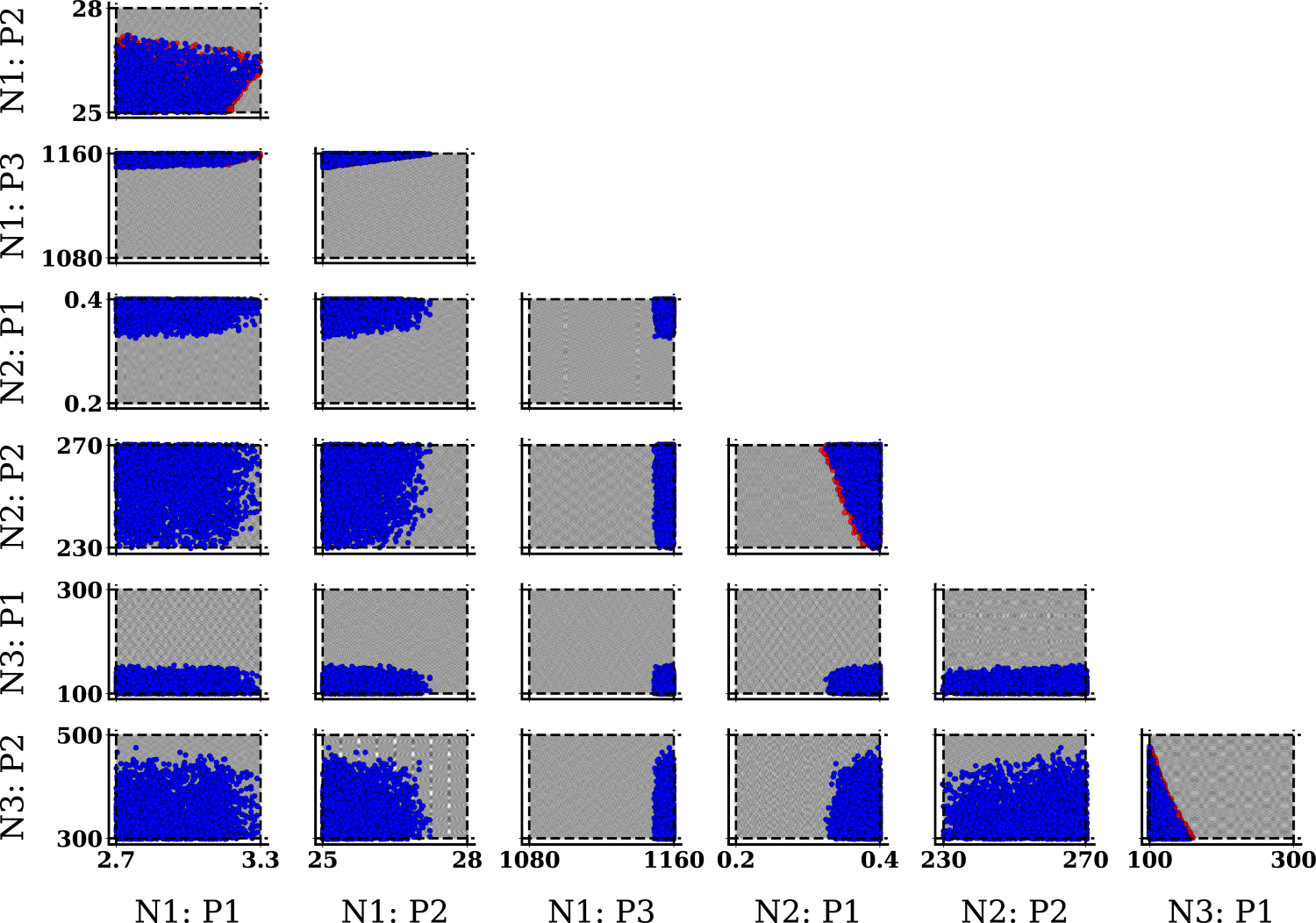}
        \caption{Results of the reconstruction.}
        \label{fig:sub2_tablet}
    \end{subfigure}
    \caption{Corner plots of the joint feasible region of operational parameters for the tablet production network at various stages of the decomposition methodology. (a) The composite backward-forward propagation results in red scatter, with samples utilized in initialization of the search domains in grey scatter. (b) The reconstruction of the joint feasible region in blue scatter, with the backward-forward propagation results in red scatter underlaid. }
    \label{fig:main_tablet}
\end{figure}

\subsubsection{Results of the Decomposition Strategy}

The following analysis investigates a composite backward-forward propagation, followed by reconstruction of the joint feasible region according to the methodology detailed in Section \ref{sec:decomp}.

The problem decomposes unit-wise via the proposed approach into three subproblems, with $1\prec 2 \prec 3$, of 3, 4 and 3 dimensions, respectively. Initialization of the search domains $\mathcal{K}^{\sf(b)}_{u_1}$, $\mathcal{K}^{\sf(b)}_{u_2}$, and $\mathcal{K}^{\sf(b)}_{u_3}$ leveraged forward simulation of the composite function representative of the process network given $8192$ Sobol samples drawn from $\mathcal{K}_v$. Outer-approximations to the node input variable domains were then identified as a box enclosure after inflating the range of the data in each dimension by 5\%.   

The subproblem ${\mathbb{VU}}^{\sf (b)}_3$ for the feed frame and tablet press is constituted by the three variables $(\textbf{v}_3,{u}_3^2)\in\mathbb{R}^2\times\mathbb{R}$. An inner-approximation to ${\mathbb{VU}}^{\sf (b)}_3$ with 3500 samples was determined by sampling. Next, the neural network classifier $\overline{\textbf{G}}^{\sf(b)}_3$ was trained to couple ${\mathbb{VU}}^{\sf (b)}_3$ to the subproblem ${\mathbb{VU}}^{\sf (b)}_2$ for the convective blender upstream defined on the four variables $(\textbf{v}_2,\textbf{u}_2^1)\in\mathbb{R}^2\times\mathbb{R}^2$. An inner-approximation to $\mathbb{VU}_2^{\sf (b)}$ was again represented using 3500 samples. The neural network classifier $\overline{\textbf{G}}^{\sf (b)}_2$ was trained to subsequently couple ${\mathbb{VU}}^{\sf (b)}_2$ to the subproblem ${\mathbb{VU}}^{\sf (b)}_1$ for the co-mill upstream, defined on the three parameters $\textbf{v}_1 \in \mathbb{R}^3$. Again, ${\mathbb{VU}}^{\sf (b)}_1$ was inner-approximated with 3500 samples identified through nested sampling, which concludes the backward propagation.

The forward propagation step started with the co-mill subproblem $\mathbb{VU}_1^{\sf (bf)} = \mathbb{VU}_1^{\sf (b)}$. The neural network classifier $\overline{\textbf{G}}^{\sf (bf)}_1$ was trained to subsequently couple $\mathbb{VU}^{\sf (bf)}_1$ to the convective blender subproblem ${\mathbb{VU}}^{\sf (bf)}_2$ downstream, along with the neural network regressor $\overline{\textbf{F}}_{1}^{2}$ to approximate the constituent function describing the co-mill. The local feasible set ${\mathbb{VU}}^{\sf (bf)}_2$ provided by nested sampling again used 3500 points, after which the neural network classifier $\overline{\textbf{G}}^{\sf(bf)}_2$ was retrained along with a neural network regressor of the node function evaluation $\overline{\textbf{F}}_{2}^{3}$. These surrogate functions were finally used to identify ${\mathbb{VU}}^{\sf (bf)}_3$. The three local feasible sets $\mathbb{V}^{\sf (bf)}_1$, $\mathbb{V}^{\sf (bf)}_2$ and $\mathbb{V}^{\sf (bf)}_3$ are visualized by red scatter in five subplots of Figure~\ref{fig:sub1_tablet}. 

The final step of reconstructing the joint feasible region relied on uniform sampling of the discrete Cartesian product set ${\mathbb{V}}^{\sf (bf)} \coloneqq {\mathbb{V}}^{\sf (bf)}_1\times {\mathbb{V}}^{\sf (bf)}_2\times {\mathbb{V}}^{\sf (bf)}_3$ without adaptation. This final result of the decomposition approach is detailed by 3500 samples on Figure~\ref{fig:sub2_tablet} in blue scatter, showing little overestimation compared to the red scatter generated from the backward-forward propagations. This indicates that the composed propagations provide a tight reduced search domain for the reconstruction in this case study. Finally, comparison of Figure~\ref{fig:directable} and Figure~\ref{fig:sub2_tablet} reveals that the simultaneous and decomposition procedures recover similar feasible regions, with differences attributable to the use of random sampling. 
\FloatBarrier

\subsection{Estimation and Verification of an Approximator}\label{sec:facs}
\subsubsection{Problem Statement}
Consider the following non-convex function ${{f}:\mathcal{K}_z \rightarrow \mathbb{R}}$,\footnote{It is worth highlighting that the indexing of scalar variables in \eqref{eq:ncvxfn} differentiates input variables to the function rather than coupling parameters lifts to a given node.} defined by
\begin{equation}\label{eq:ncvxfn}
    {f}(\textbf{z}) \coloneqq \sum_{m=1}^{n_z} (z_m^3 - z_m^2) - \sum_{m=1}^{n_z} \sum_{n>m}^{n_z} z_mz_n \,.
\end{equation}
%
%
It is desired to identify the set of parameter values ${\bf v}\in\mathcal{K}_v$ for which a parametric function $\bar{f}(\cdot, \textbf{v})$ is a valid approximation to this function on $\mathcal{K}_z$ within a given tolerance threshold, $\epsilon \in \mathcal{K}_\epsilon \subset \mathbb{R}_+$. Therefore, we seek to identify the subset
\begin{equation*}
    \mathbb{VZE} = \left\{ (\textbf{v},\textbf{z}, \epsilon) \in  \mathcal{K}_v\times \mathcal{K}_z \times \mathcal{K}_\epsilon~\middle| \begin{array}{l}
        \lvert\bar{f}(\textbf{z}; \textbf{v}) -  {f}(\textbf{z})\rvert_p \leq \epsilon
    \end{array} \right\}.
\end{equation*}

The structure of $\bar{f}$ is chosen to be the following sum of basis functions, \vspace{4mm}
\begin{equation*}
    \eqnmark[black]{node6}{\bar{f}(\textbf{z}; \textbf{v})} \coloneqq \eqnmark[black]{node5}{\textbf{z}^\intercal ({\bf L}{\bf L}^\intercal) \textbf{z}} + \eqnmark[black]{node4}{\textbf{p}^\intercal \textbf{z}} + \eqnmark[black]{node3}{\sum_{m=1}^{n_z}z_m S_m \log(z_m +1)} - \eqnmark[black]{node2}{\sum_{m=1}^{n_z}T_m \log(z_m + 1)} + \eqnmark[black]{node1}{c}.
    \annotate[yshift=1em]{left}{node6}{Node 6} 
    \annotate[yshift=-1em]{below,left}{node5}{Node 5} 
    \annotate[yshift=-1em]{below,right}{node4}{Node 4} 
    \annotate[yshift=0.5em]{right}{node3}{Node 3}
    \annotate[yshift=0.5em]{below,right}{node2}{Node 2} 
    \annotate[yshift=1em]{right}{node1}{Node 1}
\end{equation*}
\vspace{4mm}

\noindent
The corresponding connectivity in the DAG of the approximator $\bar{f}$ is described below. 
\begin{center}
    \begin{tikzpicture}
    \node[draw, circle] (1) at (-4,-2) {1};
    \node[draw, circle] (2) at (-2,-1) {2};
    \node[draw, circle] (3) at (0,0) {3};
    \node[draw, circle] (4) at (2,-1) {4};
    \node[draw, circle] (5) at (4,-2) {5};
    \node[draw, circle] (6) at (0,-2) {6};
    \draw[->] (1) -- (6);
    \draw[->] (2) -- (6);
    \draw[->] (3) -- (6);
    \draw[->] (4) -- (6);
    \draw[->] (5) -- (6);
    \end{tikzpicture}
\end{center}

The case study investigates the case with $n_z=2$ and $\mathcal{K}_z=[-0.5,1.0]\times [0.0, 0.3]$. The compact set $\mathcal{K}_v$ describing candidate parameter values for the basis functions is chosen so that $c\in[-1,1]$ (node 1), $T_m\in[0,1]$ (node 2), $S_m\in[0,1]$ (node 3), $p_m\in[-1,1]$ (node 4), and $L_{m,n}\in[-1,1]$ for $m\geq n$, $L_{m,n}=0$ otherwise. The search domain on the point-wise error is defined as  $\mathcal{K}_\epsilon=[0,0.25]$. One may sample the set $\mathbb{VZE}$ through simultaneous solution or through decomposition with parameter lifts on the coupling variables $(\textbf{z}, \epsilon) \in \mathcal{K}_z \times \mathcal{K}_\epsilon$.


%
The in-silico data generated through sampling $\mathbb{VZE}$ allows one to identify an approximate parameterisation $\overline{\textbf{G}}(\textbf{v}, \textbf{z}, \epsilon)$ through binary classification. This parameterisation may then be used for a subsequent task. Here, we identify the function approximation that yields minimium point-wise error over the domain through the solution of a semi-infinite program,
\begin{equation}
\begin{aligned}
    &\min_{(\textbf{v}, \epsilon) \in \mathcal{K}_v\times \mathcal{K}_\epsilon} ~\epsilon \\
    &\text{s.t.}\ 0 \ge \ \overline{\textbf{G}}(\textbf{v}, \textbf{z}, \epsilon), \quad \forall \textbf{z} \in \mathcal{K}_z \,.
\end{aligned}
\end{equation}
%
We compare solutions yielded by the simultaneous approach and decomposition methods. The approximation scheme used to solve this problem followed the approach developed by \citet{jungen2023libdips}.

\subsubsection{Results of the Decomposition Strategy}
The joint problem $\mathbb{VZE}$ decomposes into 6 subproblems, as detailed above. Coupling parameter lifts are made for both the variables $\textbf{z}$, constituting the function domain, as well as the relative error threshold $\epsilon$. It is worth noting that although the subproblem for node~1 describes the setting of a constant parameter, the nature of the problem requires inclusion of these auxiliary variables across all nodes. The dimensionality of each subproblem is detailed by Table~\ref{table:cs_summary}. A backward propagation is applied, with only the constraint {\sf (a)} enforced in the extension of \eqref{eq:DS_unitBP},\footnote{The implication is that a larger set is yielded than could possibly be identified, but the embedded optimisation problems become box-constrained NLPs, which can be solved efficiently using {\sf JAX} and reduce computational overhead in the propagation.} and the set ${\mathbb{VZE}}$ reconstructed thereafter from the discrete search set. Each subproblem ${\mathbb{VUZE}}_i^{\sf(b)}$ utilizes \num{2500} samples in solution. As with the simultaneous approach, \num{15000} samples are then used to represent the joint solution ${\mathbb{VZE}}$. This is a major difference between this and previous case studies, where each subproblem solution has utilized a number of samples consistent with the simultaneous identification. The decomposition starts from node 6 and proceeds backwards, identifying ${\mathbb{VUZE}}^{\sf (b)}_i$ having built a surrogate constraint $\overline{\textbf{G}}_6^{\sf (b)}$ for the purpose of node coupling. The reduced domain $\mathbb{ZE}^{\sf(b)} = {\mathbb{ZE}}^{\sf(b)}_6$ is used for defining a reduced space in the reconstruction. The set $\mathbb{VZE}$ is then reconstructed as before and a classifier built to parameterise $\mathbb{VZE}$. The semi-infinite program defined is then solved approximately to estimate and verify an approximating function on the domain, $\mathcal{K}_z$.

\begin{figure}[ht!]
    \centering
    \begin{subfigure}{0.48\linewidth} 
        \includegraphics[width=\linewidth]{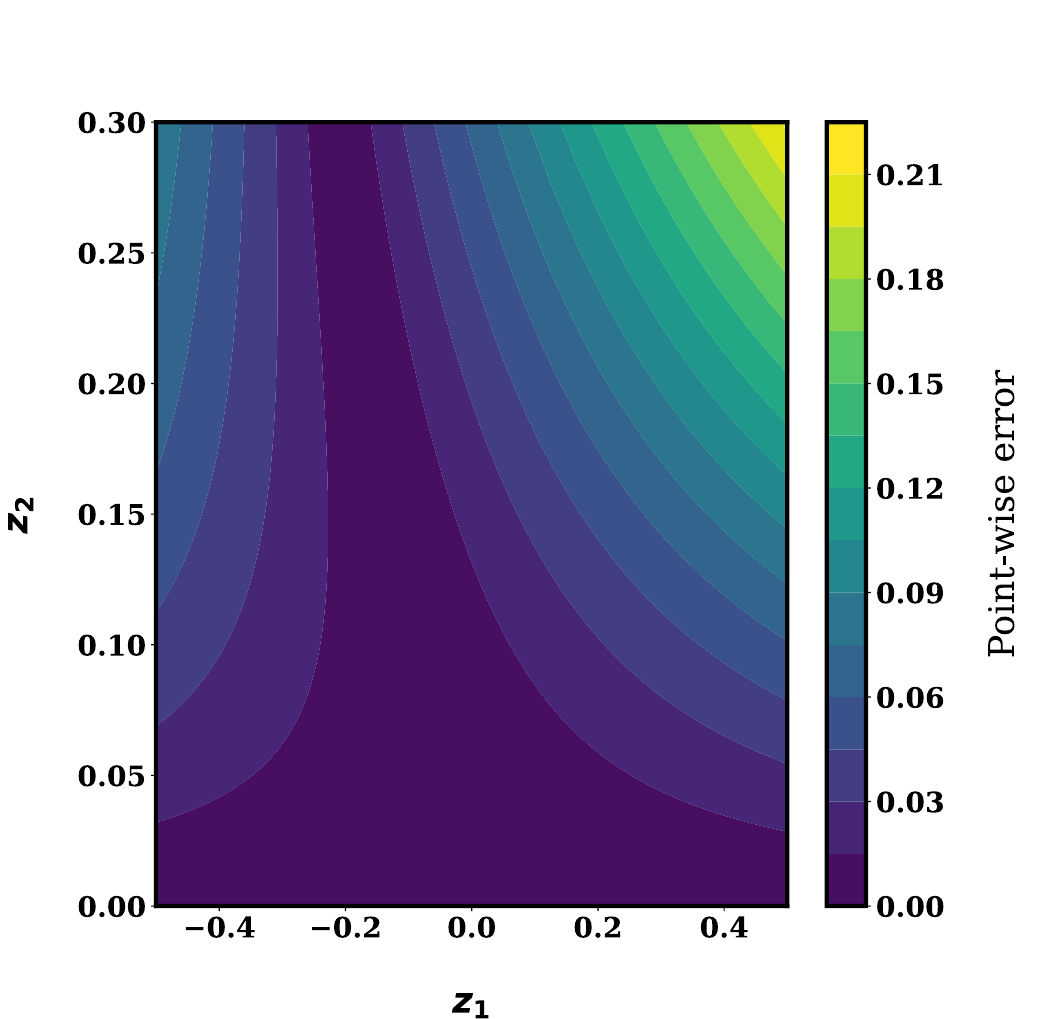}
        \caption{The simultaneous solution}
        \label{fig:subfig_d_star} 
    \end{subfigure}
    \hfill 
    \begin{subfigure}{0.48\linewidth} 
        \includegraphics[width=\linewidth]{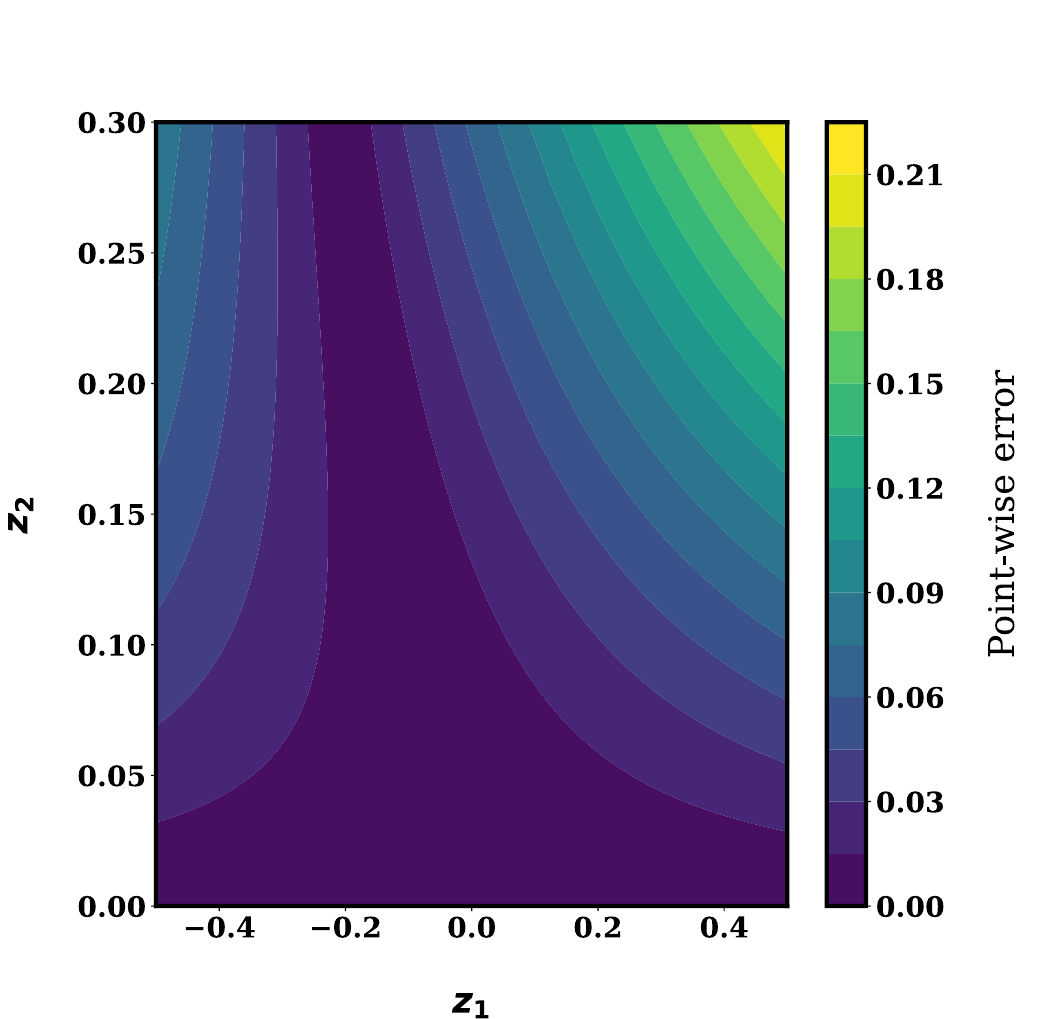}
        \caption{The decomposition solution}
        \label{fig:subfig_d_underlined} 
    \end{subfigure}
    \caption{Point-wise error surfaces for the approximators yielded by the simultaneous and decomposition approaches.}
    \label{fig:contourcomp}
\end{figure}

Figure \ref{fig:contourcomp} describes the point-wise error surface of the approximator estimated through the simultaneous and decomposition approaches. From comparison it is evident that the approximators are of similar predictive accuracy with both attaining a maximum error of approximately \num{0.21}. For the simultaneous approach the maximum predicted error of the semi-infinite program value function was \num{0.171}; whereas the decomposition scheme predicted \num{0.168}. The disparity in the solution and ground truth decomposes between approximation error in the surrogate parameterisation and the semi-infinite programming approximation scheme used. However, this study highlights that the decomposition scheme yields parameterisations of comparable quality that can be exploited in decision-making thereafter.
\FloatBarrier

\section{Results Analysis}\label{sec:CSquant}

\subsection{Accuracy of Surrogate Construction}\label{sec:accsurr}

The propagation of extended local feasible sets $\mathbb{VU}^{\sf (d)}_i$ through the nodes of the DAG entails the construction of surrogate models. In the following, we analyze both the cross-validation performance of the optimal model structure, and the training metrics of the final surrogate constructed. The analysis is provided with respect to classifiers $\overline{\textbf{G}}^{\sf(d)}_i$ trained within all propagations, as well as regressors $\overline{\textbf{F}}_i^k$ used to approximate the constituent functions within the forward propagations. 

All subproblem solutions parameterised by binary classifiers are assessed via training and validation classification accuracies {\sf (ACC)}. Additionally, all neural network classifiers are also assessed with binary cross-entropy loss {\sf (BCE)}. When no {\sf BCE} is reported, the classifier is an {SVM} model. It is fair to evaluate these metrics alone, as both classes possess the same number of datapoints due to use of a data augmentation strategy. All surrogates of the constituent functions take the form of neural network regressors trained to minimize the mean squared error {\sf (MSE)}.

\begin{table}[htb]
    \centering
    \small
    \begin{tabular}{ccccccccc}

    \toprule
         \multirow{2}{*}{\#} & \multirow{2}{*}{Propagation} & \multirow{2}{*}{Node} & \multicolumn{2}{c}{\sf BCE} & \multicolumn{2}{c}{\sf ACC} & \multicolumn{2}{c}{\sf MSE}  \\
         \cmidrule{4-9}
         &  &  &   {\sf T} & {\sf 2-CV} & {\sf T} & {\sf 2-CV} & {\sf T} & {\sf 2-CV}   \\
         \toprule
         \multirow{8}{*}{1} & \multirow{5}{*}{Forward} & 1 & 0.0015 & 0.0022 & 1.000 & 0.999 & 0.0001 & 0.0001 \\
          &  & 2 & 0.0001 & 0.0016 & 0.996 & 0.988 & 0.0001 & 0.0001 \\
          &  & 3 & 0.0159 & 0.0284 & 0.998 & 0.993 & 0.0001 & 0.0001 \\
          &  & 4 & 0.0344 & 0.0487 & 0.999 & 0.986 & 0.0001 & 0.0001 \\
         \cmidrule{2-9}
          & \multirow{4}{*}{Backward} & 5 & -- & -- & 0.995 & 0.985 & -- & -- \\
          & & 4 & -- & -- & 0.997 & 0.988 & 0.0001 & 0.0001 \\
          & & 3 & -- & -- & 0.999 & 0.987 & 0.0001 & 0.0001 \\
          & & 2 & -- & -- & 0.995 & 0.990 & 0.0001 & 0.0001 \\
         \midrule
         \multirow{1}{*}{2} & \multirow{1}{*}{Backward} & 2 & -- & -- & 1.000 & 1.000 & -- & -- \\
         \midrule
         \multirow{5}{*}{3} & \multirow{2}{*}{Backward} 
         & 3 &  0.0048 & 0.0049 & 0.998 & 0.998 & -- & --  \\
         & & 2 & 0.0534 & 0.0607 & 0.985 & 0.984 & -- & --  \\
         \cmidrule{2-9}
         & \multirow{2}{*}{Backward-Forward}  
         & 1 & 0.0061 & 0.0063 & 0.998 & 0.997 & 0.0003 &  0.0002 \\
         &  & 2 & 0.0711 & 0.090 & 0.979 & 0.981 & 0.0002 & 0.0002 \\
        \midrule
        \multirow{1}{*}{4} & \multirow{1}{*}{Backward} & 6 & -- & -- & 1.000 & 0.996 & -- & --  \\
    \bottomrule
    \end{tabular}
    \caption{Two fold cross-validation {(\sf 2-CV)} performance of the optimal classifier and regressor structures, together with final training {\sf (T)} results of the surrogate model on the whole dataset available. The metrics for classification include cross-entropy loss {\sf (BCE)} and accuracy {\sf (ACC)}, while mean squared error {\sf MSE} is used to assess the performance of regression.}
    \label{tab:my_label_class_regress}
\end{table}

The results in Table~\ref{tab:my_label_class_regress} show that in final training {\sf (T)} all neural network classifiers have a {\sf BCE} close to zero and all classifiers possess {\sf ACC} close to one across the case studies investigated. This demonstrates that binary classification frameworks provide effective parameterisations of the feasible region in the studies of interest. Additionally, all regressors were trained with low {\sf MSE}, again demonstrating the flexibility of machine learning regression frameworks. It is also worth noting that the cross-validation framework utilized was effective given the similarity between all {\sf T} and {\sf 2-CV} results reported, indicating identification of model structures that generalise well across the data distribution generated by sampling.

\subsection{Sample efficiency}\label{sec:sampeff}

Recall the working hypothesis that the decomposition methodology can improve the sampling efficiency to set-membership parameter estimation of composite functions through redefining the a priori search domain. An assessment is provided by the acceptance ratio {\sf (AR)}, given by
\begin{equation*}
    {\sf AR} = \frac{\# \ \text{Joint feasible parameter settings $(\textbf{v})$ identified}}{\# \ \text{Total constituent function evaluations}} \in [0,1]\,.
\end{equation*}
This metric essentially evaluates the number of feasible parameter settings $\textbf{v} \in \mathbb{V}$, that are identified relative to the total number of constituent function evaluations. The results are detailed in Table \ref{tab:my_table_1} for both the simultaneous and decomposition strategies. 
\begin{table}[htb]
    \centering
    \small
    \begin{tabular}{cccc}
       \# &  Method & \# $\textbf{F}_i(\textbf{v}_i, \textbf{u}_i)$ ($\times10^3$) & {\sf AR} (\%) \\
        \toprule
        \multirow{3}{*}{1} & Simultaneous & 37540  & 0.019 \\
       & Decomposition {\sf (f)} & \textbf{171} & \textbf{4.100} \\
       & Decomposition {\sf (b)} & 301 & 2.322 \\
       \midrule
      \multirow{2}{*}{2} & Simultaneous & 2881  & 0.121 \\
       & Decomposition {\sf (b)}& \textbf{391} & \textbf{0.894} \\
       \midrule
       \multirow{2}{*}{3} & Simultaneous &  8,430  & 0.042\\
       & Decomposition {\sf (bf)}& \textbf{222}  & \textbf{1.575}  \\
       \midrule
       \multirow{2}{*}{4} & Simultaneous & {198} & 0.076 \\
       & Decomposition {\sf (b)} &  \textbf{145}  & \textbf{0.103}  \\
       \bottomrule
    \end{tabular}
    \caption{Metrics of respective approaches. Assessed are the total number of constituent function evaluations (\# $\textbf{F}_i(\textbf{v}_i, \textbf{u}_i)$) to gain an approximation to $\mathbb{V}$ and the acceptance ratio ({\sf AR}). The notation `Decomposition {\sf (d)}' indicates the propagation applied.}
    \label{tab:my_table_1}
\end{table}
\paragraph{Illustrative Example ${\sf (\# 1)}$} The improvements in sample efficiency arising within the application of a forward and backward propagation respectively are significant. The simultaneous problem has 10 dimensions, while the subproblems defined have 3 dimensions across all nodes apart from node 3, which has 4 due to receiving two different input connections. In both cases, a solution constituted by 7000 parameter values for the simultaneous problem is identified with improvements in {\sf AR} of two orders of magnitude over directly sampling. In both propagations subproblems were solved using 2500 parameter values which yielded a reduced Cartesian product search set of $2500^5$ parameter values from which the solution was constructed. The forward propagation was found to be more efficient, and we posit this is due to a comparative reduction in the size of the search space on input node variables; with the forward propagation accounting for feasibility of nodes earlier in the precedence order yielding $\mathcal{K}^{\sf (f)}_{u_i}\subseteq \mathcal{K}^{\sf (b)}_{u_i}$. Now clearly, it is far more efficient to use the {\sf MPT 4.0} toolbox for solution. However, in the case one is agnostic to the underlying functions within the graph, these results demonstrate considerable potential in generalising sampling solutions to {CSP} to highly dimensional settings.

\paragraph{Serial Batch Reactor Network ${\sf (\# 2)}$} The decomposition method requires about $14\%$ the total number of constituent function evaluations in comparison to the simultaneous approach (i.e. adaptive sampling from $\mathcal{K}_{v_1}\times \mathcal{K}_{v_2}$). While case dependent, this result is encouraging insofar as the subproblem for reactor 2 in the decomposition method has the same dimensionality as the simultaneous problem itself. A likely explanation for this behavior is that the extended search domain of reactor 2, $\mathcal{K}_{vu_2}$ has a smaller volume than that of the multi-unit search region $\mathcal{K}_{v_1}\times \mathcal{K}_{v_2}$, resulting in improved acceptance rates in sampling iterations. 

\paragraph{Tableting operation ${\sf (\# 3)}$} The decomposition approach yields a two-order improvement in the {\sf AR}, requiring less than $3\%$ the total number of constituent function evaluations of the simultaneous approach. This is a substantial reduction and is a promising result in the case that the constituent functions are particularly expensive to evaluate.\footnote{It is worth noting that although the number of constituent functions evaluations is dramatically decreased through decomposition, there is requirements to solve the coupling { NLP} problems during the propagations. These are generally cheap to solve (i.e., $\sim 10~\rm ms$), but do add small overhead.}

\paragraph{Function Approximation ${\sf (\# 4)}$} The decomposition scheme required 73.2\% the number of function evaluations of the simultaneous approach. This improvement is perhaps less than expected; however, this is primarily due to the structure of the problem, as the sampler is able to determine both the left and right hand side of the constraint system. Despite the relatively small increase in the efficiency of identification, both solutions were of similar quality as evidenced by Figure \ref{fig:contourcomp}. It should be noted that in variants of this study where the point-wise error was fixed and not determined within the identification, significant improvements in {\sf AR} were often observed from decomposition due to high problem dimensionality. However, the study as presented effectively shows that functions abstracted from the decomposition scheme can be used effectively for downstream decision-making tasks.

\FloatBarrier

\section{Conclusions}

This research has developed methodology to exploit problem structure to effectively identify a set of feasible constituent function parameters for sampling-based set-membership parameter estimation of composite functions. This is a description that is flexible to many constraint satisfaction problems. 

The methodology proceeds by defining node-wise parameter estimation problems, and solves them by composing two different approaches to subproblem relaxation and propagation, i.e. backward relaxations and propagations, and forward relaxations and propagations. Issues related to defining search spaces, coupling node-wise subproblems and identifying a description of the subproblem solution itself is handled through surrogate modelling, sampling and nonlinear programming. The solutions to the node-wise subproblems are used as a reduced search space to construct the joint set of feasible parameter values. Four case studies were designed to explore the behavior of the methodology. 

The first demonstrated the approach to subproblem solution without reliance on sampling. This enabled empirical demonstration that a single relaxation and propagation identifies a superset to the node-wise subproblem. When two opposing propagations are combined, a further
tightening of the local feasible sets was observed, although the result remains an outer-approximation in general. When solved by a sampling algorithm, the forward and backward propagations enabled identification of a solution using less than 1\% of the samples of the simultaneous approach.

In the second, feasibility analysis for a network of batch reaction systems was explored. The methodology considered a more complex case where the constituent functions were evaluated as the result of a forward solution to an initial value problem. Here, the methodology was reliant on sampling. Despite the problem decomposition producing subproblems of similar dimensionality to the simultaneous approach, the methodology identified the desired joint set of feasible parameter values with a $8\times$ improvement in sampling efficiency to that of the simultaneous approach.

The third case study detailed feasibility analysis for a tableting operation. The process network was represented as a composite function of three constituent functions describing a co-mill, a convective blender, and a tablet press. Here decomposition yielded subproblems of considerably fewer dimensions than the simultaneous problem. The decomposition yielded an improvement of 2 orders of magnitude in sampling efficiency relative to the simultaneous approach. This highlights the effectiveness of the proposed approach in reducing the search space. 

The final case study explored simultaneous parameter estimation and verification of an approximating function, and demonstrated the use of auxiliary variables to reformulate the problem description for the purposes of decomposition. The simultaneous problem consisted of 13 parameters. The decomposition scheme was able to yield an approximator with the similar maximum point-wise error across the domain as the simultaneous method, demonstrating that the decomposition scheme is able not only to improve sampling efficiency but also yield parameterisation of the feasible set that may be exploited for decision-making thereafter.

The methodology proposed is flexible to different forms of constituent functions, and is amenable to non-algebraic and nonlinear evaluations. In future work, we will consider handling the presence of uncertain parameters in the constituent function description and consider methods to handle cyclic graph structure.

\ifanon
\section*{Data availability}
All data in this study are generated in silico. The code, configuration files, sampled feasible sets and trained surrogates that reproduce every result are openly available in the $\boldsymbol{\mu}\textbf{F}$ repository; the URL is withheld for review and given on the title page.

\else
\section*{CRediT authorship contribution statement}
\textbf{Max Mowbray:} Conceptualization, Methodology, Software, Investigation, Visualization, Writing -- original draft. \textbf{Nilay Shah:} Conceptualization, Supervision, Funding acquisition, Writing -- review \& editing. \textbf{Beno\^it Chachuat:} Conceptualization, Methodology, Supervision, Funding acquisition, Writing -- review \& editing.

\section*{Declaration of competing interest}
The authors declare that they have no known competing financial interests or personal relationships that could have appeared to influence the work reported in this paper.

\section*{Data availability}
All data in this study are generated in silico. The code, configuration files, sampled feasible sets and trained surrogates that reproduce every result are openly available in the $\boldsymbol{\mu}\textbf{F}$ repository at \url{https://github.com/mawbray/mu.F}.

\section*{Acknowledgements}
This work is co-funded by Eli Lilly \& Company through the Pharmaceutical Systems Engineering Lab (PharmaSEL) and by the Engineering and Physical Sciences Research Council (EPSRC) as part of its Prosperity Partnership Programme under grant EP/T518207/1. The majority of this work was developed through postdoctoral research at Imperial College London. The authors would like to thank Prof. Cleo Kontoravdi for her support throughout this project.
\fi

\appendix
\section{Illustrative Case Study Details}
The details of the function and constraint parameterisations, $\text{A}_i$ and $\text{B}_i$ $\forall i \in \mathcal{N}$ are detailed here for complete information:
\begin{equation*}
    \text{A}_1 = \begin{bmatrix}
-0.2637 & 0.2414
\end{bmatrix}; \qquad \text{B}_1 = 0.3848
\end{equation*}

\begin{equation*}
\text{A}_2 = \begin{bmatrix}
-0.7514 & 0.7332
\end{bmatrix}; \qquad \text{B}_2 = 0.2283
\end{equation*}

\begin{equation*}
    \text{A}_{3} = \begin{bmatrix}
0.7012 & -0.3727 \\
0.3304 & -0.1756 
\end{bmatrix}; \qquad \text{B}_3 = \begin{bmatrix}
    0.4949 &  -1.4861\\
    -0.6915 &  0.1617
\end{bmatrix}
\end{equation*}

\begin{equation*}
\text{A}_4 = \begin{bmatrix}
0.5356 & -0.3625
\end{bmatrix}; \qquad \text{B}_4 = 0.2442
\end{equation*}

\begin{equation*}
    \text{A}_5 = \begin{bmatrix}
-1.6635 & 0.7768
\end{bmatrix}; \qquad \text{B}_5 = 1.1443
\end{equation*}

\bibliographystyle{elsarticle-harv}
\bibliography{sample}

\begin{thebibliography}{50}
\expandafter\ifx\csname natexlab\endcsname\relax\def\natexlab#1{#1}\fi
\providecommand{\url}[1]{\texttt{#1}}
\providecommand{\href}[2]{#2}
\providecommand{\path}[1]{#1}
\providecommand{\DOIprefix}{doi:}
\providecommand{\ArXivprefix}{arXiv:}
\providecommand{\URLprefix}{URL: }
\providecommand{\Pubmedprefix}{pmid:}
\providecommand{\doi}[1]{\href{http://dx.doi.org/#1}{\path{#1}}}
\providecommand{\Pubmed}[1]{\href{pmid:#1}{\path{#1}}}
\providecommand{\bibinfo}[2]{#2}
\ifx\xfnm\relax \def\xfnm[#1]{\unskip,\space#1}\fi
\bibitem[{Albarghouthi(2021)}]{albarghouthi2021introductionneuralnetworkverification}
\bibinfo{author}{Albarghouthi, A.}, \bibinfo{year}{2021}.
\newblock \bibinfo{title}{Introduction to neural network verification}.
\newblock \URLprefix \url{https://arxiv.org/abs/2109.10317},
  \href{http://arxiv.org/abs/2109.10317}{{\tt arXiv:2109.10317}}.
\bibitem[{Andersson et~al.(2019)Andersson, Gillis, Horn, Rawlings and
  Diehl}]{Andersson2019}
\bibinfo{author}{Andersson, J.A.E.}, \bibinfo{author}{Gillis, J.},
  \bibinfo{author}{Horn, G.}, \bibinfo{author}{Rawlings, J.B.},
  \bibinfo{author}{Diehl, M.}, \bibinfo{year}{2019}.
\newblock \bibinfo{title}{{CasADi} -- {A} software framework for nonlinear
  optimization and optimal control}.
\newblock \bibinfo{journal}{Mathematical Programming Computation}
  \bibinfo{volume}{11}, \bibinfo{pages}{1--36}.
\newblock \DOIprefix\doi{10.1007/s12532-018-0139-4}.
\bibitem[{Aubin and Frankowska(1990)}]{aubin1990set}
\bibinfo{author}{Aubin, J.P.}, \bibinfo{author}{Frankowska, H.},
  \bibinfo{year}{1990}.
\newblock \bibinfo{title}{Set-valued analysis}.
\newblock \bibinfo{publisher}{Birkh\a user Boston,}.
\bibitem[{Bect et~al.(2012)Bect, Ginsbourger, Li, Picheny and
  Vazquez}]{bect2012sequential}
\bibinfo{author}{Bect, J.}, \bibinfo{author}{Ginsbourger, D.},
  \bibinfo{author}{Li, L.}, \bibinfo{author}{Picheny, V.},
  \bibinfo{author}{Vazquez, E.}, \bibinfo{year}{2012}.
\newblock \bibinfo{title}{Sequential design of computer experiments for the
  estimation of a probability of failure}.
\newblock \bibinfo{journal}{Statistics and Computing} \bibinfo{volume}{22},
  \bibinfo{pages}{773--793}.
\bibitem[{Benhamou et~al.(1999)Benhamou, Goualard, Granvilliers and
  Puget}]{benhamou1999revising}
\bibinfo{author}{Benhamou, F.}, \bibinfo{author}{Goualard, F.},
  \bibinfo{author}{Granvilliers, L.}, \bibinfo{author}{Puget, J.F.},
  \bibinfo{year}{1999}.
\newblock \bibinfo{title}{Revising hull and box consistency.}, in:
  \bibinfo{booktitle}{ICLP}, pp. \bibinfo{pages}{230--244}.
\bibitem[{Blondel et~al.(2021)Blondel, Berthet, Cuturi, Frostig, Hoyer,
  Llinares-L{\'o}pez, Pedregosa and Vert}]{jaxopt_implicit_diff}
\bibinfo{author}{Blondel, M.}, \bibinfo{author}{Berthet, Q.},
  \bibinfo{author}{Cuturi, M.}, \bibinfo{author}{Frostig, R.},
  \bibinfo{author}{Hoyer, S.}, \bibinfo{author}{Llinares-L{\'o}pez, F.},
  \bibinfo{author}{Pedregosa, F.}, \bibinfo{author}{Vert, J.P.},
  \bibinfo{year}{2021}.
\newblock \bibinfo{title}{Efficient and modular implicit differentiation}.
\newblock \bibinfo{journal}{arXiv preprint arXiv:2105.15183} .
\bibitem[{Bodirsky et~al.(2017)Bodirsky, Jonsson and
  Pham}]{bodirsky2017complexity}
\bibinfo{author}{Bodirsky, M.}, \bibinfo{author}{Jonsson, P.},
  \bibinfo{author}{Pham, T.V.}, \bibinfo{year}{2017}.
\newblock \bibinfo{title}{The complexity of phylogeny constraint satisfaction
  problems}.
\newblock \bibinfo{journal}{ACM Transactions on Computational Logic (TOCL)}
  \bibinfo{volume}{18}, \bibinfo{pages}{1--42}.
\bibitem[{Boss et~al.(2004)Boss, Elsinger, Summer and
  Thurner~4}]{boss2004network}
\bibinfo{author}{Boss, M.}, \bibinfo{author}{Elsinger, H.},
  \bibinfo{author}{Summer, M.}, \bibinfo{author}{Thurner~4, S.},
  \bibinfo{year}{2004}.
\newblock \bibinfo{title}{Network topology of the interbank market}.
\newblock \bibinfo{journal}{Quantitative finance} \bibinfo{volume}{4},
  \bibinfo{pages}{677--684}.
\bibitem[{Boukouvala and Ierapetritou(2012)}]{boukouvala2012feasibility}
\bibinfo{author}{Boukouvala, F.}, \bibinfo{author}{Ierapetritou, M.G.},
  \bibinfo{year}{2012}.
\newblock \bibinfo{title}{Feasibility analysis of black-box processes using an
  adaptive sampling kriging-based method}.
\newblock \bibinfo{journal}{Computers \& Chemical Engineering}
  \bibinfo{volume}{36}, \bibinfo{pages}{358--368}.
\bibitem[{Bradbury et~al.(2018)Bradbury, Frostig, Hawkins, Johnson, Leary,
  Maclaurin, Necula, Paszke, Vander{P}las, Wanderman-{M}ilne and
  Zhang}]{jax2018github}
\bibinfo{author}{Bradbury, J.}, \bibinfo{author}{Frostig, R.},
  \bibinfo{author}{Hawkins, P.}, \bibinfo{author}{Johnson, M.J.},
  \bibinfo{author}{Leary, C.}, \bibinfo{author}{Maclaurin, D.},
  \bibinfo{author}{Necula, G.}, \bibinfo{author}{Paszke, A.},
  \bibinfo{author}{Vander{P}las, J.}, \bibinfo{author}{Wanderman-{M}ilne, S.},
  \bibinfo{author}{Zhang, Q.}, \bibinfo{year}{2018}.
\newblock \bibinfo{title}{{JAX}: composable transformations of
  {P}ython+{N}um{P}y programs}.
\newblock \bibinfo{howpublished}{\url{http://github.com/jax-ml/jax}}.
\newblock \bibinfo{note}{Version 0.3.13}.
\bibitem[{Brailsford et~al.(1999)Brailsford, Potts and
  Smith}]{brailsford1999constraint}
\bibinfo{author}{Brailsford, S.C.}, \bibinfo{author}{Potts, C.N.},
  \bibinfo{author}{Smith, B.M.}, \bibinfo{year}{1999}.
\newblock \bibinfo{title}{Constraint satisfaction problems: Algorithms and
  applications}.
\newblock \bibinfo{journal}{European journal of operational research}
  \bibinfo{volume}{119}, \bibinfo{pages}{557--581}.
\bibitem[{Buchner(2023)}]{buchner2023nested}
\bibinfo{author}{Buchner, J.}, \bibinfo{year}{2023}.
\newblock \bibinfo{title}{Nested sampling methods}.
\newblock \bibinfo{journal}{Statistic Surveys} \bibinfo{volume}{17},
  \bibinfo{pages}{169--215}.
\bibitem[{Coffman and Graham(1972)}]{coffman1972optimal}
\bibinfo{author}{Coffman, E.G.}, \bibinfo{author}{Graham, R.L.},
  \bibinfo{year}{1972}.
\newblock \bibinfo{title}{Optimal scheduling for two-processor systems}.
\newblock \bibinfo{journal}{Acta informatica} \bibinfo{volume}{1},
  \bibinfo{pages}{200--213}.
\bibitem[{Daoutidis et~al.(2019)Daoutidis, Tang and
  Allman}]{daoutidis2019decomposition}
\bibinfo{author}{Daoutidis, P.}, \bibinfo{author}{Tang, W.},
  \bibinfo{author}{Allman, A.}, \bibinfo{year}{2019}.
\newblock \bibinfo{title}{Decomposition of control and optimization problems by
  network structure: Concepts, methods, and inspirations from biology}.
\newblock \bibinfo{journal}{AIChE Journal} \bibinfo{volume}{65},
  \bibinfo{pages}{e16708}.
\bibitem[{Echard et~al.(2011)Echard, Gayton and Lemaire}]{echard2011ak}
\bibinfo{author}{Echard, B.}, \bibinfo{author}{Gayton, N.},
  \bibinfo{author}{Lemaire, M.}, \bibinfo{year}{2011}.
\newblock \bibinfo{title}{Ak-mcs: an active learning reliability method
  combining kriging and monte carlo simulation}.
\newblock \bibinfo{journal}{Structural safety} \bibinfo{volume}{33},
  \bibinfo{pages}{145--154}.
\bibitem[{Feroz et~al.(2019)Feroz, Hobson, Cameron and Pettitt}]{Feroz_2019}
\bibinfo{author}{Feroz, F.}, \bibinfo{author}{Hobson, M.P.},
  \bibinfo{author}{Cameron, E.}, \bibinfo{author}{Pettitt, A.N.},
  \bibinfo{year}{2019}.
\newblock \bibinfo{title}{Importance nested sampling and the multinest
  algorithm}.
\newblock \bibinfo{journal}{The Open Journal of Astrophysics}
  \bibinfo{volume}{2}.
\newblock \URLprefix \url{http://dx.doi.org/10.21105/astro.1306.2144},
  \DOIprefix\doi{10.21105/astro.1306.2144}.
\bibitem[{Floudas et~al.(2001)Floudas, G{\"u}m{\"u}{\c{s}} and
  Ierapetritou}]{floudas2001global}
\bibinfo{author}{Floudas, C.A.}, \bibinfo{author}{G{\"u}m{\"u}{\c{s}}, Z.H.},
  \bibinfo{author}{Ierapetritou, M.G.}, \bibinfo{year}{2001}.
\newblock \bibinfo{title}{Global optimization in design under uncertainty:
  feasibility test and flexibility index problems}.
\newblock \bibinfo{journal}{Industrial \& Engineering Chemistry Research}
  \bibinfo{volume}{40}, \bibinfo{pages}{4267--4282}.
\bibitem[{Geremia et~al.(2024)Geremia, Bezzo and
  Ierapetritou}]{geremia2024design}
\bibinfo{author}{Geremia, M.}, \bibinfo{author}{Bezzo, F.},
  \bibinfo{author}{Ierapetritou, M.G.}, \bibinfo{year}{2024}.
\newblock \bibinfo{title}{Design space determination of pharmaceutical
  processes: Effects of control strategies and uncertainty}.
\newblock \bibinfo{journal}{European Journal of Pharmaceutics and
  Biopharmaceutics} \bibinfo{volume}{194}, \bibinfo{pages}{159--169}.
\bibitem[{Hagberg et~al.(2008)Hagberg, Swart and Schult}]{hagberg2008exploring}
\bibinfo{author}{Hagberg, A.}, \bibinfo{author}{Swart, P.J.},
  \bibinfo{author}{Schult, D.A.}, \bibinfo{year}{2008}.
\newblock \bibinfo{title}{Exploring network structure, dynamics, and function
  using NetworkX}.
\newblock \bibinfo{type}{Technical Report}. Los Alamos National Laboratory
  (LANL), Los Alamos, NM (United States).
\bibitem[{Harwood and Barton(2017)}]{harwood2017solve}
\bibinfo{author}{Harwood, S.M.}, \bibinfo{author}{Barton, P.I.},
  \bibinfo{year}{2017}.
\newblock \bibinfo{title}{How to solve a design centering problem}.
\newblock \bibinfo{journal}{Mathematical Methods of Operations Research}
  \bibinfo{volume}{86}, \bibinfo{pages}{215--254}.
\bibitem[{Jalving et~al.(2019)Jalving, Cao and Zavala}]{jalving2019graph}
\bibinfo{author}{Jalving, J.}, \bibinfo{author}{Cao, Y.},
  \bibinfo{author}{Zavala, V.M.}, \bibinfo{year}{2019}.
\newblock \bibinfo{title}{Graph-based modeling and simulation of complex
  systems}.
\newblock \bibinfo{journal}{Computers \& Chemical Engineering}
  \bibinfo{volume}{125}, \bibinfo{pages}{134--154}.
\bibitem[{Jungen et~al.(2023)Jungen, Zingler, Djelassi and
  Mitsos}]{jungen2023libdips}
\bibinfo{author}{Jungen, D.}, \bibinfo{author}{Zingler, A.},
  \bibinfo{author}{Djelassi, H.}, \bibinfo{author}{Mitsos, A.},
  \bibinfo{year}{2023}.
\newblock \bibinfo{title}{libdips--discretization-based semi-infinite and
  bilevel programming solvers}.
\newblock \bibinfo{journal}{dynamics} \bibinfo{volume}{20},
  \bibinfo{pages}{42}.
\bibitem[{Kucherenko et~al.(2020)Kucherenko, Giamalakis, Shah and
  Garc{\'\i}a-Mu{\~n}oz}]{kucherenko2020computationally}
\bibinfo{author}{Kucherenko, S.}, \bibinfo{author}{Giamalakis, D.},
  \bibinfo{author}{Shah, N.}, \bibinfo{author}{Garc{\'\i}a-Mu{\~n}oz, S.},
  \bibinfo{year}{2020}.
\newblock \bibinfo{title}{Computationally efficient identification of
  probabilistic design spaces through application of metamodeling and adaptive
  sampling}.
\newblock \bibinfo{journal}{Computers \& Chemical Engineering}
  \bibinfo{volume}{132}, \bibinfo{pages}{106608}.
\bibitem[{Kudva et~al.(2024)Kudva, Tang and Paulson}]{kudva2024robust}
\bibinfo{author}{Kudva, A.}, \bibinfo{author}{Tang, W.T.},
  \bibinfo{author}{Paulson, J.A.}, \bibinfo{year}{2024}.
\newblock \bibinfo{title}{Robust bayesian optimization for flexibility analysis
  of expensive simulation-based models with rigorous uncertainty bounds}.
\newblock \bibinfo{journal}{Computers \& Chemical Engineering}
  \bibinfo{volume}{181}, \bibinfo{pages}{108515}.
\bibitem[{Kumar(1992)}]{kumar1992algorithms}
\bibinfo{author}{Kumar, V.}, \bibinfo{year}{1992}.
\newblock \bibinfo{title}{Algorithms for constraint-satisfaction problems: A
  survey}.
\newblock \bibinfo{journal}{AI magazine} \bibinfo{volume}{13},
  \bibinfo{pages}{32--32}.
\bibitem[{Kusumo et~al.(2019)Kusumo, Gomoescu, Paulen, Garc{\'\i}a~Mu{\~n}oz,
  Pantelides, Shah and Chachuat}]{kusumo2019bayesian}
\bibinfo{author}{Kusumo, K.P.}, \bibinfo{author}{Gomoescu, L.},
  \bibinfo{author}{Paulen, R.}, \bibinfo{author}{Garc{\'\i}a~Mu{\~n}oz, S.},
  \bibinfo{author}{Pantelides, C.C.}, \bibinfo{author}{Shah, N.},
  \bibinfo{author}{Chachuat, B.}, \bibinfo{year}{2019}.
\newblock \bibinfo{title}{Bayesian approach to probabilistic design space
  characterization: A nested sampling strategy}.
\newblock \bibinfo{journal}{Industrial \& Engineering Chemistry Research}
  \bibinfo{volume}{59}, \bibinfo{pages}{2396--2408}.
\bibitem[{Kusumo et~al.(2021)Kusumo, Morrissey, Mitchell, Shah and
  Chachuat}]{kusumo2021design}
\bibinfo{author}{Kusumo, K.P.}, \bibinfo{author}{Morrissey, J.},
  \bibinfo{author}{Mitchell, H.}, \bibinfo{author}{Shah, N.},
  \bibinfo{author}{Chachuat, B.}, \bibinfo{year}{2021}.
\newblock \bibinfo{title}{A design centering methodology for probabilistic
  design space}.
\newblock \bibinfo{journal}{IFAC-PapersOnLine} \bibinfo{volume}{54},
  \bibinfo{pages}{79--84}.
\bibitem[{Kvasnica et~al.(2004)Kvasnica, Grieder and Baoti\'{c}}]{mpt}
\bibinfo{author}{Kvasnica, M.}, \bibinfo{author}{Grieder, P.},
  \bibinfo{author}{Baoti\'{c}, M.}, \bibinfo{year}{2004}.
\newblock \bibinfo{title}{{Multi-Parametric Toolbox (MPT)}}.
\newblock \URLprefix \url{http://control.ee.ethz.ch/~mpt/}.
\bibitem[{Mart{\'\i}nez-Jaramillo et~al.(2010)Mart{\'\i}nez-Jaramillo,
  P{\'e}rez, Embriz and Dey}]{martinez2010systemic}
\bibinfo{author}{Mart{\'\i}nez-Jaramillo, S.}, \bibinfo{author}{P{\'e}rez,
  O.P.}, \bibinfo{author}{Embriz, F.A.}, \bibinfo{author}{Dey, F.L.G.},
  \bibinfo{year}{2010}.
\newblock \bibinfo{title}{Systemic risk, financial contagion and financial
  fragility}.
\newblock \bibinfo{journal}{Journal of Economic Dynamics and Control}
  \bibinfo{volume}{34}, \bibinfo{pages}{2358--2374}.
\bibitem[{Mitsos et~al.(2008)Mitsos, Lemonidis and Barton}]{mitsos2008global}
\bibinfo{author}{Mitsos, A.}, \bibinfo{author}{Lemonidis, P.},
  \bibinfo{author}{Barton, P.I.}, \bibinfo{year}{2008}.
\newblock \bibinfo{title}{Global solution of bilevel programs with a nonconvex
  inner program}.
\newblock \bibinfo{journal}{Journal of Global Optimization}
  \bibinfo{volume}{42}, \bibinfo{pages}{475--513}.
\bibitem[{Mohr and Henderson(1986)}]{mohr1986arc}
\bibinfo{author}{Mohr, R.}, \bibinfo{author}{Henderson, T.C.},
  \bibinfo{year}{1986}.
\newblock \bibinfo{title}{Arc and path consistency revisited}.
\newblock \bibinfo{journal}{Artificial intelligence} \bibinfo{volume}{28},
  \bibinfo{pages}{225--233}.
\bibitem[{Moritz et~al.(2018)Moritz, Nishihara, Wang, Tumanov, Liaw, Liang,
  Elibol, Yang, Paul, Jordan et~al.}]{moritz2018ray}
\bibinfo{author}{Moritz, P.}, \bibinfo{author}{Nishihara, R.},
  \bibinfo{author}{Wang, S.}, \bibinfo{author}{Tumanov, A.},
  \bibinfo{author}{Liaw, R.}, \bibinfo{author}{Liang, E.},
  \bibinfo{author}{Elibol, M.}, \bibinfo{author}{Yang, Z.},
  \bibinfo{author}{Paul, W.}, \bibinfo{author}{Jordan, M.I.}, et~al.,
  \bibinfo{year}{2018}.
\newblock \bibinfo{title}{Ray: A distributed framework for emerging $\{$AI$\}$
  applications}, in: \bibinfo{booktitle}{13th USENIX symposium on operating
  systems design and implementation (OSDI 18)}, pp. \bibinfo{pages}{561--577}.
\bibitem[{Ochoa et~al.(2021)Ochoa, Garc{\'\i}a-Mu{\~n}oz, Stamatis and
  Grossmann}]{ochoa2021novel}
\bibinfo{author}{Ochoa, M.P.}, \bibinfo{author}{Garc{\'\i}a-Mu{\~n}oz, S.},
  \bibinfo{author}{Stamatis, S.}, \bibinfo{author}{Grossmann, I.E.},
  \bibinfo{year}{2021}.
\newblock \bibinfo{title}{Novel flexibility index formulations for the
  selection of the operating range within a design space}.
\newblock \bibinfo{journal}{Computers \& Chemical Engineering}
  \bibinfo{volume}{149}, \bibinfo{pages}{107284}.
\bibitem[{Paulen et~al.(2015)Paulen, Villanueva and Chachuat}]{paulen2015gpe}
\bibinfo{author}{Paulen, R.}, \bibinfo{author}{Villanueva, M.E.},
  \bibinfo{author}{Chachuat, B.}, \bibinfo{year}{2015}.
\newblock \bibinfo{title}{Guaranteed parameter estimation of non-linear dynamic
  systems using high-order bounding techniques with domain and cpu-time
  reduction strategies}.
\newblock \bibinfo{journal}{IMA Journal of Mathematical Control \& Information}
  \bibinfo{volume}{33}, \bibinfo{pages}{563--587}.
\newblock \DOIprefix\doi{10.1093/imamci/dnu055}.
\bibitem[{Picheny et~al.(2010)Picheny, Ginsbourger, Roustant, Haftka and
  Kim}]{picheny2010adaptive}
\bibinfo{author}{Picheny, V.}, \bibinfo{author}{Ginsbourger, D.},
  \bibinfo{author}{Roustant, O.}, \bibinfo{author}{Haftka, R.T.},
  \bibinfo{author}{Kim, N.H.}, \bibinfo{year}{2010}.
\newblock \bibinfo{title}{Adaptive designs of experiments for accurate
  approximation of a target region}.
\newblock \bibinfo{journal}{Journal of Mechanical Design}
  \bibinfo{volume}{132}, \bibinfo{pages}{071008}.
\newblock \DOIprefix\doi{10.1115/1.4001873}.
\bibitem[{Pulsipher and Zavala(2019)}]{pulsipher2019scalable}
\bibinfo{author}{Pulsipher, J.L.}, \bibinfo{author}{Zavala, V.M.},
  \bibinfo{year}{2019}.
\newblock \bibinfo{title}{A scalable stochastic programming approach for the
  design of flexible systems}.
\newblock \bibinfo{journal}{Computers \& Chemical Engineering}
  \bibinfo{volume}{128}, \bibinfo{pages}{69--76}.
\bibitem[{Puranik and Sahinidis(2017)}]{puranik2017domain}
\bibinfo{author}{Puranik, Y.}, \bibinfo{author}{Sahinidis, N.V.},
  \bibinfo{year}{2017}.
\newblock \bibinfo{title}{Domain reduction techniques for global nlp and minlp
  optimization}.
\newblock \bibinfo{journal}{Constraints} \bibinfo{volume}{22},
  \bibinfo{pages}{338--376}.
\bibitem[{Ravanbakhsh and Greiner(2015)}]{ravanbakhsh2015perturbed}
\bibinfo{author}{Ravanbakhsh, S.}, \bibinfo{author}{Greiner, R.},
  \bibinfo{year}{2015}.
\newblock \bibinfo{title}{Perturbed message passing for constraint satisfaction
  problems}.
\newblock \bibinfo{journal}{The Journal of Machine Learning Research}
  \bibinfo{volume}{16}, \bibinfo{pages}{1249--1274}.
\bibitem[{del Rio-Chanona et~al.(2016)del Rio-Chanona, Fiorelli and
  Vassiliadis}]{del2016automated}
\bibinfo{author}{del Rio-Chanona, E.A.}, \bibinfo{author}{Fiorelli, F.},
  \bibinfo{author}{Vassiliadis, V.S.}, \bibinfo{year}{2016}.
\newblock \bibinfo{title}{Automated structure detection for distributed process
  optimization}.
\newblock \bibinfo{journal}{Computers \& Chemical Engineering}
  \bibinfo{volume}{89}, \bibinfo{pages}{135--148}.
\bibitem[{Rockafellar and Wets(1998)}]{rockafellar1998set}
\bibinfo{author}{Rockafellar, R.T.}, \bibinfo{author}{Wets, R.J.},
  \bibinfo{year}{1998}.
\newblock \bibinfo{title}{Set convergence}.
\newblock \bibinfo{journal}{Variational analysis} , \bibinfo{pages}{108--147}.
\bibitem[{Sachio et~al.(2023)Sachio, Kontoravdi and
  Papathanasiou}]{sachio2023model}
\bibinfo{author}{Sachio, S.}, \bibinfo{author}{Kontoravdi, C.},
  \bibinfo{author}{Papathanasiou, M.M.}, \bibinfo{year}{2023}.
\newblock \bibinfo{title}{A model-based approach towards accelerated process
  development: A case study on chromatography}.
\newblock \bibinfo{journal}{Chemical Engineering Research and Design}
  \bibinfo{volume}{197}, \bibinfo{pages}{800--820}.
\bibitem[{Schichl and Neumaier(2005)}]{schichl2005interval}
\bibinfo{author}{Schichl, H.}, \bibinfo{author}{Neumaier, A.},
  \bibinfo{year}{2005}.
\newblock \bibinfo{title}{Interval analysis on directed acyclic graphs for
  global optimization}.
\newblock \bibinfo{journal}{Journal of Global Optimization}
  \bibinfo{volume}{33}, \bibinfo{pages}{541--562}.
\bibitem[{Smith(2005)}]{smith2005chemical}
\bibinfo{author}{Smith, R.}, \bibinfo{year}{2005}.
\newblock \bibinfo{title}{Chemical process: design and integration}.
\newblock \bibinfo{publisher}{John Wiley \& Sons}.
\bibitem[{Song et~al.(2022)Song, Cao, Zhang, Xu and Lim}]{song2022learning}
\bibinfo{author}{Song, W.}, \bibinfo{author}{Cao, Z.}, \bibinfo{author}{Zhang,
  J.}, \bibinfo{author}{Xu, C.}, \bibinfo{author}{Lim, A.},
  \bibinfo{year}{2022}.
\newblock \bibinfo{title}{Learning variable ordering heuristics for solving
  constraint satisfaction problems}.
\newblock \bibinfo{journal}{Engineering Applications of Artificial
  Intelligence} \bibinfo{volume}{109}, \bibinfo{pages}{104603}.
\bibitem[{Swaney and Grossmann(1985)}]{swaney1985index_I}
\bibinfo{author}{Swaney, R.E.}, \bibinfo{author}{Grossmann, I.E.},
  \bibinfo{year}{1985}.
\newblock \bibinfo{title}{An index for operational flexibility in chemical
  process design. {Part I}: Formulation and theory}.
\newblock \bibinfo{journal}{AIChE Journal} \bibinfo{volume}{31},
  \bibinfo{pages}{621--630}.
\bibitem[{Van~Beek(2006)}]{van2006backtracking}
\bibinfo{author}{Van~Beek, P.}, \bibinfo{year}{2006}.
\newblock \bibinfo{title}{Backtracking search algorithms}, in:
  \bibinfo{booktitle}{Foundations of artificial intelligence}.
  \bibinfo{publisher}{Elsevier}. volume~\bibinfo{volume}{2}, pp.
  \bibinfo{pages}{85--134}.
\bibitem[{W{\"a}chter and Biegler(2006)}]{wachter2006implementation}
\bibinfo{author}{W{\"a}chter, A.}, \bibinfo{author}{Biegler, L.T.},
  \bibinfo{year}{2006}.
\newblock \bibinfo{title}{On the implementation of an interior-point filter
  line-search algorithm for large-scale nonlinear programming}.
\newblock \bibinfo{journal}{Mathematical programming} \bibinfo{volume}{106},
  \bibinfo{pages}{25--57}.
\bibitem[{Yadan(2019)}]{Yadan2019Hydra}
\bibinfo{author}{Yadan, O.}, \bibinfo{year}{2019}.
\newblock \bibinfo{title}{Hydra - a framework for elegantly configuring complex
  applications}.
\newblock \bibinfo{howpublished}{Github}.
\newblock \URLprefix \url{https://github.com/facebookresearch/hydra}.
\bibitem[{Zhou and Peng(2020)}]{zhou2020structural}
\bibinfo{author}{Zhou, T.}, \bibinfo{author}{Peng, Y.}, \bibinfo{year}{2020}.
\newblock \bibinfo{title}{Structural reliability analysis via dimension
  reduction, adaptive sampling, and monte carlo simulation}.
\newblock \bibinfo{journal}{Structural and Multidisciplinary Optimization}
  \bibinfo{volume}{62}, \bibinfo{pages}{2629--2651}.
\bibitem[{Zhu et~al.(1997)Zhu, Byrd, Lu and Nocedal}]{zhu1997algorithm}
\bibinfo{author}{Zhu, C.}, \bibinfo{author}{Byrd, R.H.}, \bibinfo{author}{Lu,
  P.}, \bibinfo{author}{Nocedal, J.}, \bibinfo{year}{1997}.
\newblock \bibinfo{title}{Algorithm 778: L-bfgs-b: Fortran subroutines for
  large-scale bound-constrained optimization}.
\newblock \bibinfo{journal}{ACM Transactions on mathematical software (TOMS)}
  \bibinfo{volume}{23}, \bibinfo{pages}{550--560}.

\end{thebibliography}

\end{document}